 \documentclass[journal]{IEEEtran}
\usepackage{cite}
\usepackage{amsmath,amssymb,amsfonts}
\usepackage{algorithmic}
\usepackage{textcomp}

\usepackage{dcolumn}% Align table columns on decimal point
\usepackage{bm}% bold math
\usepackage{epsfig}
\usepackage{subcaption}
\usepackage{longtable,booktabs,array}

\usepackage{etoolbox}
\usepackage{soul}

\makeatletter
\patchcmd\longtable{\par}{\if@noskipsec\mbox{}\fi\par}{}{}
\makeatother

\def\BibTeX{{\rm B\kern-.05em{\sc i\kern-.025em b}\kern-.08em
    T\kern-.1667em\lower.7ex\hbox{E}\kern-.125emX}}
\begin{document}
%\history{Date of publication xxxx 00, 0000, date of current version xxxx 00, 0000.}
%\doi{10.1109/ACCESS.2017.DOI}

\title{Interdependency and cascading failures in co-patenting and shareholding interfirm networks}
%
%\author{
%\uppercase{Daniel Marcolin}\authorrefmark{1},
%and
%\uppercase{Yasuyuki Todo}\authorrefmark{2},
%and
%\uppercase{Mahendra Piraveenan}\authorrefmark{1},  
%}
%
%\address[1]{Modelling and Simulation Group, School of Computer Science, Faculty of Engineering, The University of Sydney,  NSW 2006, Australia}
%
%\address[2]{Graduate School of Economics, Waseda University,  Sinjuku City, Tokyo, Japan}

\author{\IEEEauthorblockN{1\textsuperscript{st} Daniel Marcolin}\\
\IEEEauthorblockA{  \textit{School of Computer Science, Faculty of Engineering} \\ 
\textit{University of Sydney, Sydney, Australia}\\
 }
\and 
 \IEEEauthorblockN{2\textsuperscript{nd} Yasuyuki Todo}\\
\IEEEauthorblockA{\textit{Graduate School of Economics}\\ 
\textit{Waseda University,  Sinjuku City, Tokyo, Japan}\\
}
\and 
\IEEEauthorblockN{3\textsuperscript{rd} Mahendra Piraveenan} \\
\IEEEauthorblockA{\textit{School of Computer Science, Faculty of Engineering}\\ 
\textit{University of Sydney, Sydney, Australia}\\
\textit{Email: Mahendrarajah.Piraveenan@sydney.edu.au}\\
0000-0001-6550-5358}
}

%\markboth
%{Piraveenan \headeretal: Information theoretic characterisation of interfirm networks}
%{Piraveenan \headeretal:  Information theoretic characterisation of interfirm networks}
%
%\corresp{Corresponding author: Mahendra Piraveenan (e-mail: mahendrarajah.piraveenan@sydney.edu.au).}

\maketitle

\begin{abstract}

This work analyses the interdependent link creation of patent and shareholding links in interfirm networks, and how this dynamics affects the resilience of such networks in the face of cascading failures.      Using the Orbis dataset, we construct very large co-patenting and shareholding networks, globally as well as in terms of individual countries.  Besides, we construct  smaller overlap networks from those firm pairs which have both types of links between them, for nine years between 2008-2016.       We use information theoretic measures,  such as mutual information, active information storage, and transfer entropy, to characterise the topological similarities and shared topological information between the relevant  co-patenting and shareholding networks. We then construct a cascading failure model, and use it to analyse the resilience of  interdependent interfirm networks in terms of multiple failure characteristics. We find that there is relatively high level of mutual information between co-patenting networks and the shareholding networks from later years, suggesting that the formation of shareholding links is influenced by the existence of patent links in previous years.       We highlight that this phenomena differs between countries.  For interfirm networks from certain countries, such as Switzerland and Netherlands, this influence is remarkably higher compared to other countries.        We also show that this influence becomes most apparent after a delay of four years between the formation of  co-patenting links and shareholding links. Analysing the resilience of shareholding networks against cascading failures, we show that in terms of both mean downtime, and failure proportion of firms, certain countries including Italy, Germany, India,  Japan and the United States,  have less resilient shareholding networks compared to other countries with significant economies. Based on our results,  we postulate that an interfirm network model which considers multiple types of relationships together,  uses information theoretic measures to establish information sharing and causality between them,  and uses cascading failure simulation to understand the resilience of such networks under economic and financial stress, could be a useful multifaceted model to highlight important features of economic systems around the world.

\end{abstract}


%\titlepgskip=-15pt

\section{Introduction} \label{introduction}

The study of interfirm networks is an important means of characterising and examining relationships between firms    ~\cite{sydow1998organizing,  koka2006evolution, lorenzoni1999leveraging, batonda2003approaches, meira2010management,  piraveenan2019assortativity, bell2017network}. Relationships between firms are complex, and can be analysed across various domains. However, studying such relationships at a national or global level typically requires working with a large dataset. Such datasets have been used in the context of studying international supply-chain relationships, leading to insights about their structure and operation~\cite{piraveenan2019assortativity, piraveenan2020topology}.  Examining the evolution of interfirm networks by looking at time series data has also proved fruitful, with previous work in this vein showing, for instance, that Japanese customer-supplier networks demonstrate relatively low levels of link-switching each year~\cite{mizuno2015buyer}.

The current work takes a similar approach, but with a focus instead on interfirm patent collaboration relationships and shareholding relationships.  This work also seeks to understand how the topology of such networks could contribute to cascading firm failures in the event of an economic shock. In the shareholding network, nodes represent firms, and an edge is formed when a firm directly owns shares in another firm.   The network is constructed as a directed graph, with edges running from company to shareholder, since this is the direction of any potential cascading failure. On the other hand, in a co-patenting network, nodes represent firms, and an edge between firms  is formed when a firm co-owns a patent with another firm. Therefore a co-patenting network is undirected.  This analysis is carried out using the Orbis database, a private database compiled by analytics company Bureau van Dijk, which contains data on over 400 million companies~\cite{orbisdata}.

Both  co-patenting networks (which are also, in a sense, research collaboration networks since the patents result from research collaborations between firms)  and shareholding networks have been shown to provide effective lenses through which we can analyse aspects of the economy. For example, the structure of interfirm collaboration networks may play a key role in the innovative output of firms~\cite{schilling2007interfirm},  whereas the structure of interfirm shareholding networks may play a role in the spread of economic shocks~\cite{dastkhan2019simulation}.  This paper aims to determine whether there is a relationship between the evolution of these networks.

Specifically, this work uses information-theoretic measures to investigate whether edge formation in one network provides any information about edge formation in the other, and whether there is any directionality to such a relationship. Information-theoretic measures have been used in the study of complex networks to quantify aspects of those networks~\cite{sole2004information},  but in this case methods are adopted to enable such measures to be applied to the study of edge formation between different networks. When two firms collaborate on a patent application, this forms an edge in the patent collaboration network. In the case of the shareholding network, any shareholding interest held by one firm in another firm (whether directly or indirectly via an interposed company) forms an edge between those firms. If there is a connection between these two processes, this could provide a broader perspective on interfirm relationships, including the time frame over which they develop from one type of relationship to another. The results show some evidence that the existence of patent links provides information about the existence of future shareholding links, rather than the other way around.       A more fine-grained analysis is also undertaken to investigate variations in these results based on whether the links are intra-national or international.  We also analyse variations on a country-by-country basis, and some possible reasons for these differences are discussed.

This paper also investigates cascading failures in interfirm networks, particularly shareholding networks.Typical cascading failure models (which consider some form of `load' which may be redirected away from failed nodes, causing other nodes to `overload') are not easily applicable to  shareholding networks, since there is not a good analogue for this `load' in shareholding networks. Therefore, we first develop and implement a cascading failure model which reflects the basic properties of the shareholding network and can be applied to a very large dataset. We then investigate what factors contribute to the severity of cascading firm failures, including whether there are any significant differences between countries based on their shareholding nettwork topology. In general terms, the approach taken is to first simulate an economic shock in which a percentage of the total nodes are removed from the network. The simulation then advances over a number of time steps, during which neighbouring nodes (i.e., shareholders of the failed firms) also have a chance to fail. The failure probabilities are calculated based on the model parameters, and actual firm failures are computed stochastically based on these failure probabilities. The effects of multiple neighbours failing may accumulate, but firms are also able to adapt to losses of their investments (i.e., shareholdings) over time.  The presented results  highlight the factors that affect the tolerance or resilience of interfirm networks to cascading failures.

The rest of the paper is organised as follows. Section \ref{background} discusses relevant previous work. Section \ref{methodology} describes the Orbis dataset, the general topological properties of shareholding and co-patenting networks obtainable from it, and the details of simulation experiments undertaken. Section \ref{results} describe the results of the simulation experiments, in terms of the causality of interdependent edge formation (co-evolution) of shareholding and co-patenting networks, as well as resilience of shareholding networks to cascading failures. Section \ref{conclusions} summarises key observations, and indicates directions for future work.

\section{Background} \label{background}

This section summarises relevant  existing literature to provide context for this work. This includes prior work relating to inter-firm networks generally, their role in contributing to economic benefits and risks, shareholding networks specifically, and cascading failure. 

Inter-firm networks have been studied from various points of interest, using approaches from industrial economics and organisational research, as well as sociological and socio-psychological approaches    ~\cite{grandori1995inter, ness2005evolution}   . It is useful to first provide a definition of interfirm networks. Grandori and Soda~\cite{grandori1995inter} consider inter-firm networks to be: \textit{A mode of regulating interdependence between firms which is different from the aggregation of these units within a single firm and from coordination through market signals (prices, strategic moves, tacit collusion, etc.)} [original emphasis]. In reviewing the extensive prior literature on inter-firm networks, they also draw a distinction between equity networks (which we call shareholding networks) and non-equity networks, with the latter category including franchising networks, joint ventures, sub-contracting networks and overlapping directorships. Ozman~\cite{ozman2009inter} adds to this  second category, networks that are formed by informal relations, mergers, acquisitions, Research and Development alliances, know-how trading and licensing arrangements. We could also include supply-chain networks, which are becoming increasingly interconnected~\cite{piraveenan2019assortativity, piraveenan2020topology}.

Interconnectedness between firms may have benefits for individual firms and for the economy. Firms are the main actors in the process of innovation, much of which takes place via the dynamics on inter-firm networks. Therefore, inter-firm collaboration may play an important role in increasing innovative output~\cite{schilling2007interfirm}. However, interconnectedness can also lead to the propagation of systemic risk, and increase the likelihood of cascading firm failures. As Heath et al. ~\cite{heath2016ccps} note, the structure of inter-firm networks can affect the severity of economic crises. They point out that in the wake of the global financial crisis, regulators moved to central clearing of over-the-counter derivatives in order to reduce interconnectedness. Huang et al. ~\cite{huang2013cascading} use modelling to demonstrate how systemic risk and the potential for cascading firm failure can result from interconnectedness between firms in the banking system.

Although shareholding networks have been studied in the literature as a specific type of inter-firm network, this is often from the perspective of seeking to understand the dynamics of ownership and control within an economy. For example, Mizuno and Kurizaki ~\cite{mizuno2020network} sought to use  shareholding networks to measure the influence of investors on companies using a network power index. Rotundo and D'Arcangelis ~\cite{rotundo2010ownership} also examined ownership and control (in this case specifically amongst Italian banks and insurance companies), but also examined the diversification of shareholdings. They found that shareholdings followed a power-law distribution, with most companies having low diversification (although large-cap companies tended to have more diversified holdings than small-cap companies). There is less literature that focusses specifically on the potential for cascading failures in shareholding networks, although Dastkhan and Gharneh ~\cite{dastkhan2019simulation} do provide some evidence that the structure of shareholding networks may play a role in the spread of economic shocks.

Typical cascading failure models, such as those who are used in electric power networks, do not seem to be directly applicable in the context of inter-firm shareholding networks.  Guo et al.  ~\cite{guo2017critical} provide a summary of the various approaches taken in the literature to modelling cascading failures in power systems. In these models, a failure of components causes a redistribution of power, which then leads to the overloading of other network components, which leads to cascading failures. However, there is no analogous `load' to be redistributed in the case of shareholding networks. Therefore in this work we  instead develop an alternative probabilistic model for cascading failure in shareholding networks, which reflects the underlying properties of those specific networks.

\section{Methodology}  \label{methodology}

\subsection{Network formation}

\subsubsection{co-patenting network}

The data available from the  Orbis database~\cite{orbisdata} to form the co-patenting network relates to patent applications made between 1895 and 2016 (inclusive). There are 107,311 unique companies in the database which held patents at the time of analysis, thus forming 107,311 nodes in the co-patenting network. Since the same companies may collaborate on multiple patent applications, it is possible to form a multigraph by considering each application as a separate link between companies. Based on this approach, the network has 2,542,637 links, forming a very large complex network. Alternatively, a network could be constructed as a simple non-weighted graph by taking the existence of at least one link between companies as an unweighted link between these companies. Based on that approach, there are 222,312 unique (firm-to-firm)  links in the network.

\begin{table}[htbp]
\caption{Topological Metrics for the co-patenting network}
\begin{center}
\begin{tabular}{|c|c|}
\hline
Metric &  Number  \\
\hline
Nodes: &  107,311 \\
Edges (multigraph) & 2,542,637 \\
Edges (Simple graph) & 222,312  \\
Average Degree (Multigraph): & 47.39   \\
Average Degree (Simple graph): &  4.14  \\
Median Degree (Multigraph): &  52.47  \\
Median Degree (Simple graph): &  5. 38  \\
Connected Components: &  15,647   \\
Nodes in Largest Connected Component: & 70,995 (66.16\%)    \\
\hline
\end{tabular}
\label{table1}
\end{center}
\end{table}

\begin{figure}[htbp]
\centering
		\includegraphics[scale = 0.3]{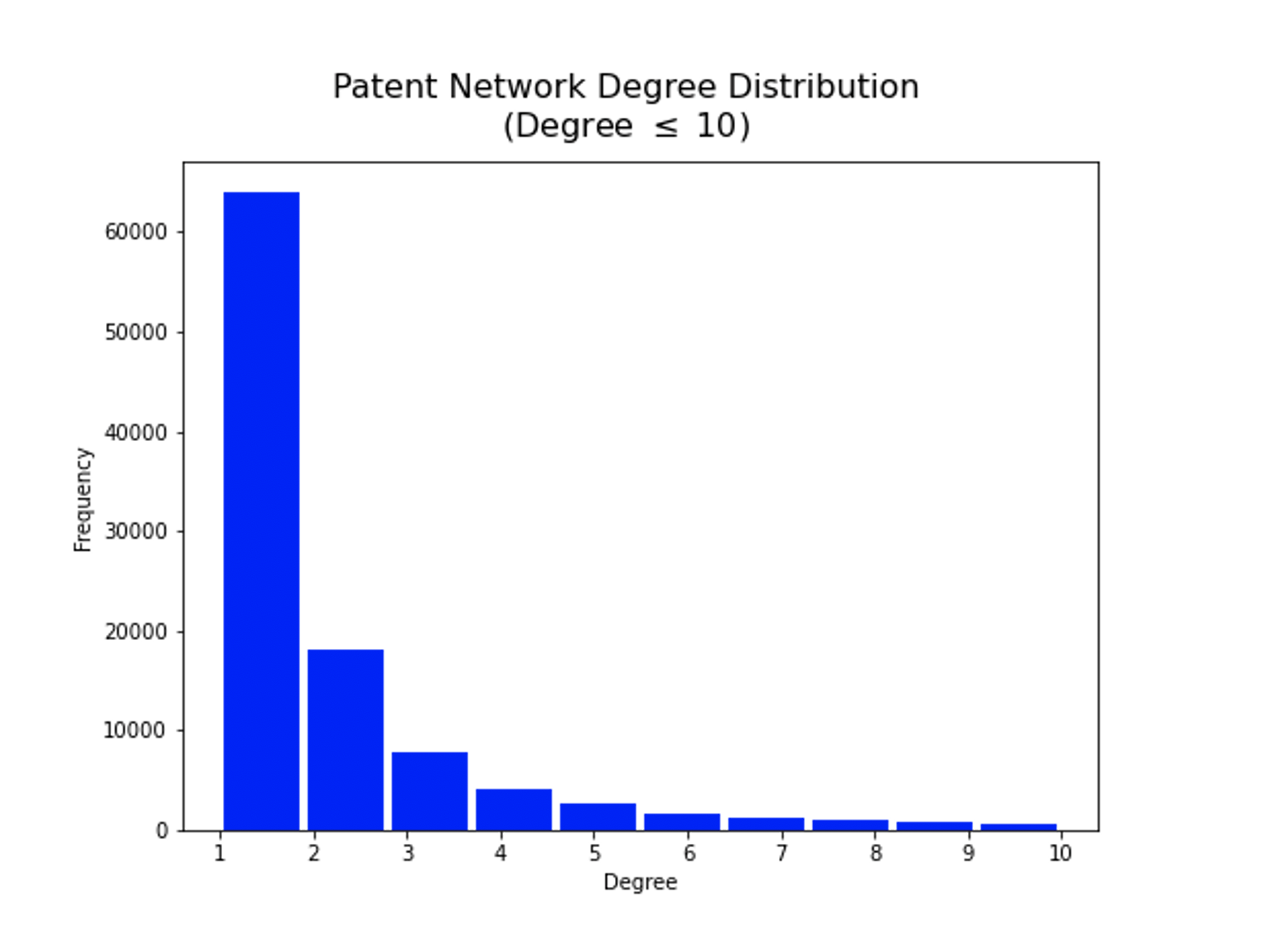}
\caption{Degree distribution of the co-patenting network. Only degrees less than ten are shown in the figure.  The distribution corresponds to the simple graph. } 
\label{fig1A}
\end{figure}

These  and other network metrics, namely the average  and median degree, the number of connected components and the number (and percentage) of nodes in the largest connected component, are set out in Table \ref{table1}. The degree distribution (for degree $\le$ 10) is shown in Fig. \ref{fig1A}. The large difference in edges between the multigraph and the simple graph, the large number of connected components, and the large number of nodes with degree 1 all suggest that there is a strong tendency for companies to repeatedly collaborate with a single research partner.

\subsubsection{Shareholding Network}

There are some 54,176,893 firms which held shares in another firm  according data available from the  Orbis database~\cite{orbisdata}. There were 42,255,702 shareholding links between them, which we have treated as unweighted (the value of shares represented was not considered). Therefore the shareholding network is not highly connected; nevertheless  the shareholding network is an extremely large complex network, and indeed would be among the largest complex networks analysed in any domain of complex network science~\cite{boccaletti2006complex, costa2007characterization, strogatz2001exploring, bell1997transportation, kasthurirathna2014influence, uddin2016set}. Unlike the co-patenting network, the shareholding network is directed.

The data available to form the shareholding network from the Orbis database~\cite{orbisdata} covers only the period between 2008 and 2017 (inclusive). However, the size of the shareholding network, it could be noted,  is much larger than the co-patenting network, with over 54 million nodes and over 42 million edges. It should be noted that in this network, links are also formed between companies with indirect shareholding interests. Thus, if a company A owns shares in another company B, which in turn owns shares in a third company C, there would be an edge between A and C, in addition to edges A--B and B--C. This is likely one reason why the network is so large.

\begin{figure}[htbp]
\centering
		\includegraphics[scale = 0.35]{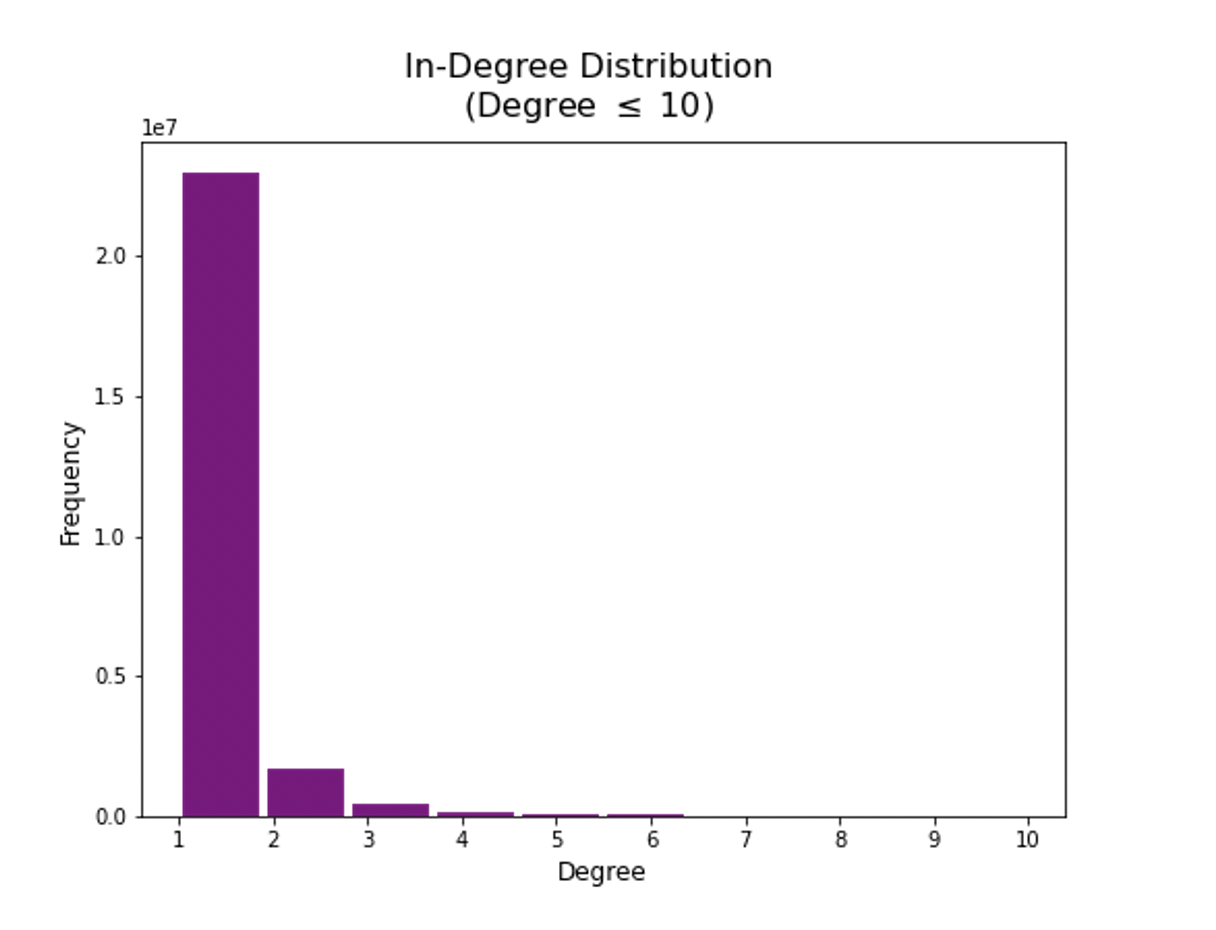}
\caption{The In-degree distribution of the shareholding network.  The shareholding network is directed, and has an extremely similar out-degree distribution. } 
\label{fig1B}
\end{figure}

Table  \ref{table2} contains some basic topological measures for the  shareholding network, and  the in-degree distribution  of the network (for degree $\le$ 10) is shown in Fig. \ref{fig1B}.  We specifically focus on the in-degree distribution in this case, because the network is directed, and the in-degrees will determine the incoming influence of economic shocks from neighbours to the node in concern.  It can be seen that the degree distribution is even more heterogenous than the  degree distribution of the co-patenting network, with a large number of nodes having degree of 1 and a relatively small number of nodes forming the largest connected component. The out-degree distribution, not shown, is extremely similar. This suggests that the network is dominated by isolated parent-subsidiary relationships, which are not part of a larger global network of interconnected shareholding relationships.

It should be mentioned that the co-patenting and shareholding networks were not constructed longitudinally: that is, despite the availability of information in the Orbis dataset about the year in which each link was created, all links which were in existence at the time of analysis were considered together to create single co-patenting and shareholding networks.  Therefore, the co-patent network represents all co-patent relationships that existed in 2016, and the shareholding network represents all shareholdinng links that existed in 2017, which are the latest years respectively in the Orbis dataset for each type of relationship. This is in contrast to the overlap network creation which we will describe next, where networks were created longitudinally: that is, links which were in existence at each year were considered separately, and separate overlap networks were created for each year that we have considered.  The reason for doing this is to create a `time-series' of overlap network topologies which will be needed for information theoretic analysis, as we will describe in the following sections.

\begin{table}[htbp]
\caption{Topological metrics for the  shareholding network}
\begin{center}
\begin{tabular}{|c|c|}
\hline
Metric &  Number  \\
\hline
Nodes: &  54,176,893  \\
Edges  & 42,255,702  \\
Average Degree: & 1.56  \\
Median Degree: & 2.11  \\
Connected Components: &  15,106,842  \\
Nodes in Largest Connected Component: & 8,219,665 (15.70\%)    \\
\hline
\end{tabular}
\label{table2}
\end{center}
\end{table}

\subsubsection{Overlap network}

We now describe the creation of the `overlap networks' between the co-patenting network and shareholding network. To compare edge formation in the patent and shareholding networks, it is first necessary to limit enquiry to the overlapping timespan for which there is data for both networks in the Orbis dataset~\cite{orbisdata}. Since the shareholding data begins in 2008 and the patent data ends in 2016, the creation of overlap networks is based on analysis conducted for the period between 2008 and 2016 (inclusive), constituting 9 years of data. 

\begin{figure*}[htbp]
\centering
		\includegraphics[scale = 0.3]{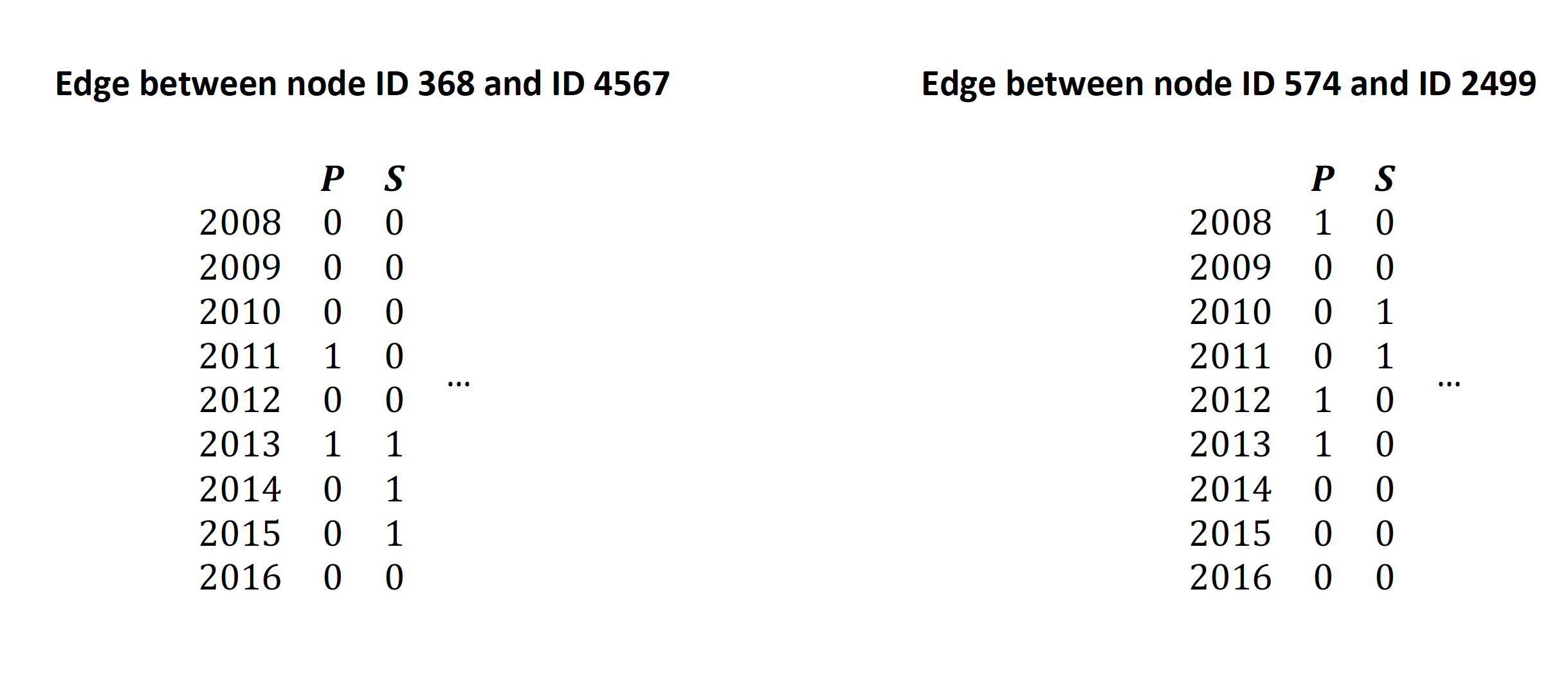}
\caption{Sample edge-existence matrices, for two particular pairs of nodes. In each case,  the column $P$ represents co-patenting links, and the column $S$ represents  shareholding links. } 
\label{fig3C}
\end{figure*}

In order to quantify edge formation in the two networks, a binary edge existence matrix is first created for each potential edge in the overlap network (each possible pair of nodes), with rows corresponding to the time series information and columns representing the existence of an edge in the co-patenting or shareholding network. In the case of the co-patenting network, a 1 indicates that a patent application was made by the two companies in the given year, whereas a 0 indicates that no such application was made. In the case of the shareholding network, a 1 indicates that a shareholding relationship was in existence  at the end of the given year, whereas a 0 indicates that no such relationship was in existence. For illustrative purposes, example edge existence matrices for two edges are shown in Fig. \ref{fig3C}, one for an edge between nodes 368 and  4567, and another for an edge between nodes  574 and 2499.

%
%\textit {Edge between node ID 368 and ID 4567}
%
%\(\begin{matrix}
% & P & S \\
%2008 & 0 & 0 \\
%2009 & 0 & 0 \\
%2010 & 0 & 0 \\
%2011 & 1 & 0 \\
%2012 & 0 & 0 \\
%2013 & 1 & 1 \\
%2014 & 0 & 1 \\
%2015 & 0 & 1 \\
%2016 & 0 & 0 \\
%\end{matrix}\) \ldots{} 
%
%\textit  {Edge between node ID 574 and ID 2499}
%
%\(\begin{matrix}
% & P & S \\
%2008 & 1 & 0 \\
%2009 & 0 & 0 \\
%2010 & 0 & 1 \\
%2011 & 0 & 1 \\
%2012 & 1 & 0 \\
%2013 & 1 & 0 \\
%2014 & 0 & 0 \\
%2015 & 0 & 0 \\
%2016 & 0 & 0 \\
%\end{matrix}\) \ldots{}

\begin{figure}[htbp]
\centering
		\includegraphics[scale = 0.3]{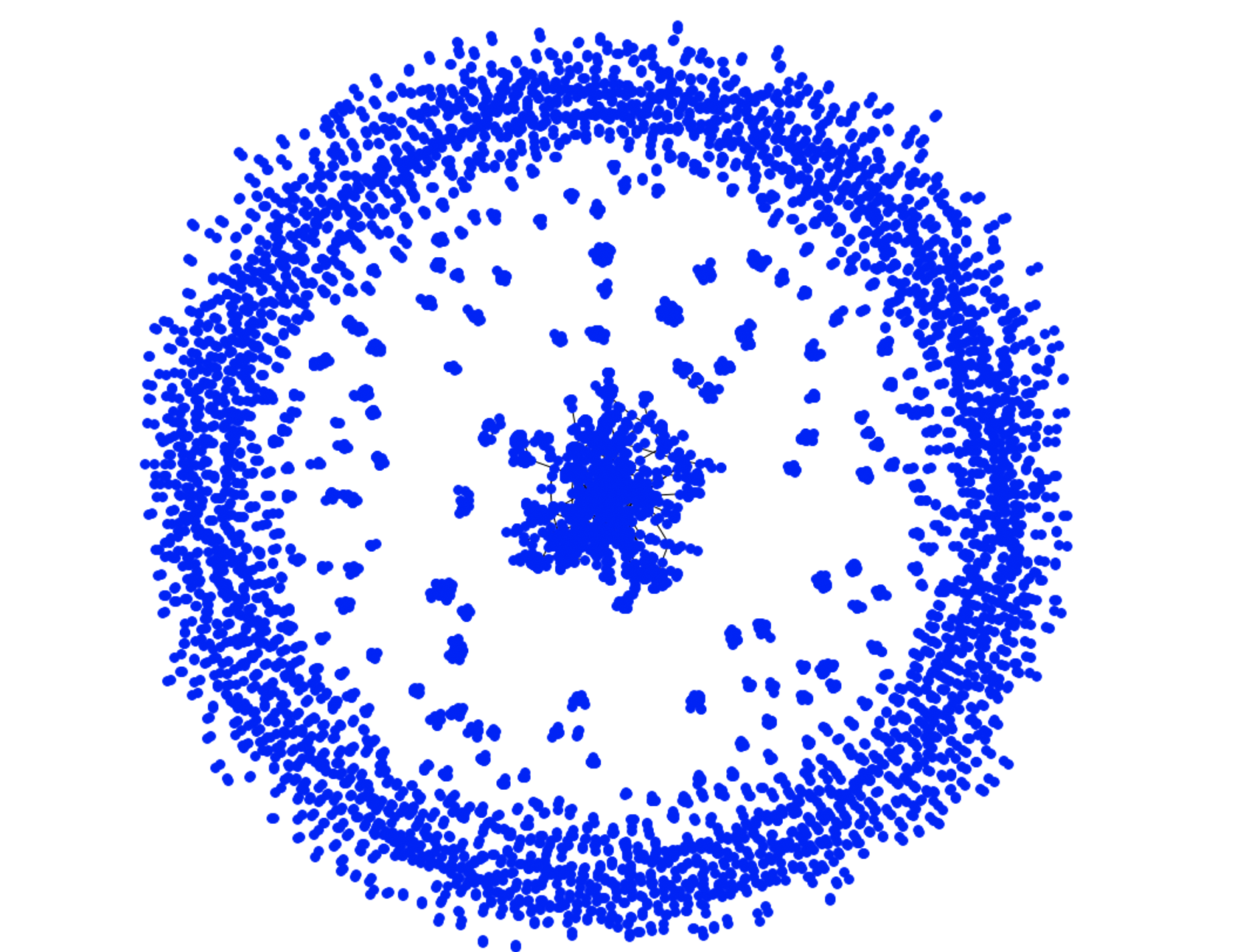}
\caption{The overlap network of the shareholding network and the co-patenting network as at 2016.} 
\label{fig4A}
\end{figure}

\begin{figure}[htbp]
\centering
		\includegraphics[scale = 0.28]{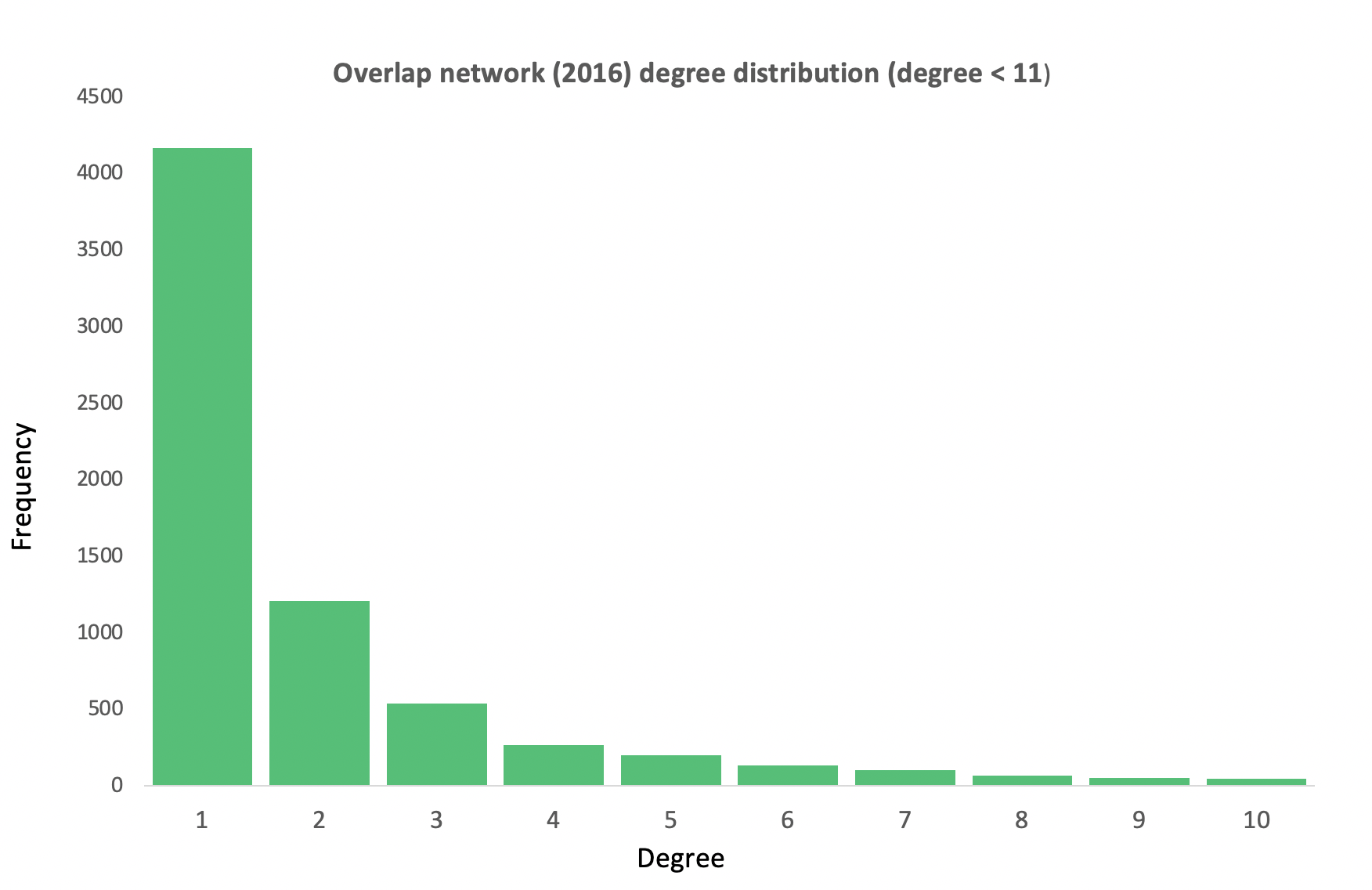}
\caption{Degree distribution of the overlap network considered (as at 2016) that is shown in Fig. \ref{fig4A}} 
\label{fig3A}
\end{figure}

Once the edge existence matrices were completed, they were used to create the overlap network.  The  overlap network  for a particular year would have a link between  nodes $v_x$ and $v_y$  only if both  co-patenting ($P$) and shareholding ($S$) columns of the corresponding edge-existence matrix were `1'  for that year. There are only a relatively small number of pairs of companies with both shareholding and co-patenting links  at a given year: hence the overlap network between the shareholding and co-patenting networks for a given year  is a small subnetwork of both the co-patenting network and shareholding network. This overlap network for 2016, for instance, had 6,842 nodes and 5,149 edges.  A visual representation of the overlap network for 2016 is shown in Fig. \ref{fig4A}. Topological metrics for this particular overlap network are shown in Table \ref{table3}, and the degree distribution (for degree $\le$ 10) is shown in Fig. \ref{fig3A}. 

\begin{table}[htbp]
\caption{Topological metrics  of the overlap network as at 2016}
\begin{center}
\begin{tabular}{|c|c|}
\hline
Metric &  Number  \\
\hline
Nodes: &  6,842   \\
Edges  & 5,149  \\
Average Degree: & 1.51  \\
Nodes in Largest Connected Component: & 1,214 (17.74\%)    \\
\hline
\end{tabular}
\label{table3}
\end{center}
\end{table}

\subsection{Information-Theoretic Analysis}

Given the overlap network  and the edge existence matrices as described above,  we then used  information-theoretic measures  to examine the causal relationship between edge formation in the two networks. The measures used for this enquiry are the mutual information and transfer entropy, with varying amounts of delay from the source to the target variable. Active information storage was also calculated for each variable, in order to determine the appropriate target embedding length to use for the transfer entropy analysis.

We, therefore, first describe the relevant information theoretic measures that we use.

\noindent
\textbf{Mutual information} is defined as follows~\cite{thomas2006elements}:

\begin{equation} \label{eq3}
I(X;Y) = \sum_{y \in Y}^{}{\sum_{x \in X}^{}{P_{(X,Y)}(x,y)\log\left( \frac{P_{(X,Y)}(x,y)}{P_{X}(x)P_{Y}(y)} \right)}}
\end{equation} where \emph{X} and \emph{Y} are a pair of random variables, \(P_{(X,Y)}\) denotes the joint distribution of \emph{X} and \emph{Y};
and \(P_{X}\) and \(P_{Y}\) denote the marginal distributions of \emph{X}
and \emph{Y} respectively.

Intuitively, this quantifies the amount of information that one random variable contains about another random variable. If the two random variables are independent, then their mutual information is 0. If on the other hand, there is a strong relationship between the random variables, a higher mutual information would be expected. It is a symmetric and nonnegative measure, and thus will provide the same result regardless of  directionality (in this case, which of the co-patenting and shareholding data is treated as the source or target variable). However, it is possible to consider the mutual information between one variable and a time-delayed set of values for  the other  variable (for example, co-patenting links from 2008 and shareholding links from 2016),  which could help give hints about causality, and this is an approach taken in our experiments. As mutual information quantifies the amount of Shannon information, it is measured in units such as bits.

\noindent
\textbf{Active information storage}, on the other hand, quantifies the amount of information about a variable that is contained in its past states. It is defined as the mutual information between the next state of a variable and its history of length k, and is calculated as follows~\cite{lizier2012local}: 

\begin{equation} \label{eq2}
A_{X}(k) = I\left( X_{n}^{(k)};X_{n + 1} \right)
\end{equation} where \(k\) denotes the target embedding length, \(X_{n + 1}\) denotes the next state of random variable \emph{X}, and \(X_{n}^{(k)}\) denotes the history of length \emph{k} of random variable \emph{X}.

\noindent
\textbf{Transfer entropy}      was introduced by Schreiber~\cite{schreiber2000measuring} as the deviation from independence of the state transition of an information destination X from the previous state of an information source Y. It is defined therefore for  two random variables   $X_n$ and $Y_n$  as ~\cite{lizier2008local, schreiber2000measuring}:

\begin{equation} \label{eq2.2}
T_{	Y \rightarrow X}  =  \sum_{u_n}^{}{{p(q_n)\log\frac{p(x_{n+1} |  x^k_{n},  y^l_{n} )}{p(x_{n+1} |  x^k_{n} )} }}
\end{equation} where  $n$ is a time index, $q_n$ is the state transition tuple $(x_{n+1},  x^k_{n}, y^l_{n}   )$, and $x^k_{n}$, $y^l_{n}$ represent  the $k$ and $l$ past values of variables $X$ and $Y$ from timestep  $n$. It can be shown  ~\cite{lizier2008local, marschinski2002analysing} that transfer entropy  is the conditional mutual information between a source variable and target variable, with the condition being the target variable's history~\cite{lizier2008local, schreiber2000measuring}.   Therefore, it can alternatively be defined using Shannon entropy as  ~\cite{marschinski2002analysing}:

\begin{equation} \label{eq2.1}
T_{	Y \rightarrow X}\left( k,l \right)  = H\left( X_{n} |  X_{n-1 : n-k}  \right) - H\left( X_{n} |  X_{n-1 : n-k},  Y_{n-1 : n-l}   \right)
\end{equation} where $k$ denotes the target history length, and $l$ denotes the source history length. Thus, it is intuitively a measure of `information transfer' between one variable and another, and may provide an indication of the directionality of the relationship.

Transfer entropy can take a number of additional parameters, including the sampling intervals for $X$ and $Y$, which can be denoted as  \(\tau_{X}\)  and  \(\tau_{Y}\), and a time delay between source to target, which we denote as $u$. Standard definitions of transfer entropy assume that the sampling intervals are equal to unity ~\cite{lizier2008local, marschinski2002analysing}.  Therefore the more general form of transfer entropy which includes all these parameters is ~\cite{marschinski2002analysing}:     .

\begin{equation} \label{eq3}
T_{Y \rightarrow X}\left( k,l,\tau_{X},\tau_{Y},u \right) = I\left( Y_{n + 1 - u}^{\left( l,\tau_{Y} \right)};X_{n + 1} \middle| X_{n}^{\left( k,\tau_{X} \right)} \right)
\end{equation} where      \(k\) denotes the target history length, \(l\) denotes the source history length, \(\tau_{X}\) denotes the target sampling interval, \(\tau_{Y}\) denotes the source sampling interval, \(u\) denotes the delay from source to target   , \(Y_{n + 1 - u}^{\left( l,\tau_{Y} \right)}\) denotes the next state of the source variable, with embedding length \emph{l} measured at intervals of \(\tau_{Y}\), and delayed by \(u\) time steps, \(X_{n + 1}\)denotes the next state of the target variable, and \(X_{n}^{\left( k,\tau_{X} \right)}\)denotes the history of the target variable, of length \emph{k} and measured at intervals of \(\tau_{X}\).

    Schreiber ~\cite{schreiber2000measuring} proposes that the `most natural choices' for the source history length $l$ are $l=k$ or $l=1$. Noting that transfer entropy measures the state transition of the target variable  from the previous state of the source variable, and is essentially conditional mutual information (with the condition being the target variable's (not source variable's) history),  if there is no context-specific requirement, there would be no need to choose a source history length other than unity. Therefore we choose  $l=1$ in this work. Given that we look at network data for each year, the sampling intervals  are also chosen as one (year).  We do not have the data to sample more frequently, and sampling less frequently would result in non-optimal use of the available data. Thus, $\tau_{X} =  \tau_{Y} = 1$.       For the purposes of the transfer entropy analysis between patent and shareholding networks, the target history (embedding) length $k$ has been set in order to maximise active information storage.        This resulted in the choice of target history length (embedding length) $k=5$ being used in the transfer entropy calculation, as this was the target embedding length that resulted in maximum active information storage, as shown in section \ref{results}. The source-to-target delay $u$ was varied, and used as a parameter in simulation experiments, as described in section \ref{results}.      

     Therefore, to summarise, the parameters used in the transfer entropy calculation are $l=1, k=5,  \tau_{X} = 1,  \tau_{Y} = 1$ and the source-target delay $u$ was varied between experiments.

Experiments were carried out using  the Java Information Dynamics Toolkit (JIDT~\cite{lizier2014jidt}), a toolkit for computing information-theoretic measurements.  Measurements were taken for each edge in the overlap network, and then averages were computed across  all edges. P-values are calculated for each edge. Since the measurements carried out on each edge can be considered  as repeated tests of the same hypothesis, the p-values are then combined using Fisher's method     ~\cite{elston1991fisher, dai2014modified}    to yield a combined p-value for each experiment.

\subsection{Cascading failure modelling}

\subsubsection{Model Overview}

We simulated cascading failures in the shareholding network, to quantify  the resilience of this network  under cascading economic events. The cascading failure model computes cascading failures stochastically using the adjacency matrix $\bm{A}$ of the network and the in-degree vector $\bm{d_{w}}$ of the network.   The results of the simulation are recorded in a `failure matrix' $\bm{F}$. The following definitions are relevant. \\

\noindent
$\bm{A}$  - the network adjacency matrix with size $N \times N$, where $N$ is the number of nodes of the network \\
\\
$\bm{d_{w}}$ -  the in-degree vector of length $N$, where each element $d_{w}^i$ represents the in-degree of node $i$, obtained by $A1_N^T$, where $1_N$  is a vector of 1s of length $N$. \\
\\
$\bm{F}$ - the failure matrix with size $T \times N$, which contains a 1 for each element $F_{(t,i)}$ where node $i$ was down during time step $t.$ Initialised as matrix $0_{(T \times N)}$. \\

At each time step, the following two vectors are dynamically updated: \\

\noindent$\bm{p_t}$ - the probability vector, where each element $p_{(t,i)}$ is the probability that node $i$ will fail during time step $t$. Initialised as $0_N$,  which is a vector of 0s of length $N$.\\
$\bm{c_t}$ - the cascade vector, which records a 1 for each element $c_{(t,i)}$ if node $i$ failed during time step $t$. Initialised as $0_N$, which is a vector of 0s of length $N$.  \\

Each row of $F$ represents a time step of the simulation, and each column represents a node in the network. Therefore, the contents of $F$ can be represented as a series of stacked vectors:

\begin{equation} \label{eq4}
F = \begin{bmatrix}
f_{1} \\
f_{2} \\
\ldots \\
f_{T} \\
\end{bmatrix}
\end{equation}

where each row $f_t$ contains the results of the simulation for time step $t$.

\subsubsection{Model Parameters}

The following parameters are used in the model. Given that the simulation is run over timesteps $T$:

\begin{itemize}

\item{$\bm{k}$ -  the failure rate, which determines the probability that a node will fail, given that a particular neighbour of it failed in  the  previous time step.  This is essentially the `transmission rate' of node failure, similar to the transmission rate of infection that is used in epidemiology~\cite{yeung2023agent}.}

\item{$\bm{r}$ -  the discount rate, which determines the rate at which the contribution to the failure probability from  a particular neighbour which failed in previous time steps diminishes over time.  In other words, it is the rate of change of the transmission rate of failure. This is because a node which has undergone a shock event is most likely to cause a neighbour to fail immediately after that shock, and this likelihood of causing failure to a neighbour is expected to decrease over time: $r$ represents the rate of this decrease}

\item{$\bm{p_i}$ -  The node failure probability of node $i$.}

\item{$\bm{\alpha}$ -  the cumulative failure rate, which determines the probability that a node will fail, given that its neighbour failed in  $any$  previous time step.  In other words, this is the failure rate when the simulation is assumed to have been completed in one time-step.}

\item{$\bm{\gamma}$ -  the  overall discount  rate, which determines the amount by which the contribution to the failure probability from  a neighbour which failed in previous time steps diminishes over time.   In other words, this is the discount rate when the simulation is assumed to have been completed in one time-step. }

\end{itemize}

The rationale for the  existence of parameter $r$ is that if a firm fails, its shareholder may have a relatively high chance to fail in the next few time steps. However, as time goes on, the shareholder will have the ability to adapt (for example, by cutting costs or finding new revenue streams). Therefore, the shareholder will have a lower probability to fail as time goes on.

Both of the cumulative parameters \(\alpha\) and \(\gamma\)  are defined with respect to a model with
\(T = 1\) (that is, assuming that the entire simulation is modelled over
a single time step).  For a model with multiple timesteps, \emph{k} is calculated in such a way that a failure rate
\emph{k} in each of the \emph{T} time steps gives the same
 failure  rate as \(\alpha\) over a single time step:

\begin{equation} \label{eq5}
k = 1 - (1 - \alpha)^{\frac{2}{T + 1}}
\end{equation}

Similarly, \emph{r} is calculated in such a way that a failure probability
diminishing by \emph{r} over \emph{T} time steps results in the same
failure probability as if it had diminished by a rate of \(\gamma\)
continuously compounded over a single time step:

\begin{equation} \label{eq6}
r = e^{\frac{\gamma}{T}} - 1
\end{equation}

The failure probability of each node $x_i$  is updated each time step. The
failure probability should be proportional to \emph{k}, but inversely
proportional to  the node in-degree ${d_{w}^i}$. This is because if a shareholder has many
shareholdings, it is less likely that failure of one of them would cause
the shareholder to fail. The failure probability should also therefore
be proportional to the number of failed shareholdings. This however means that the failure probability for a node 
is undefined for any node with in-degree zero \(d_{w}^{i} = 0\). For such cases,  we explicitly define the corresponding failure probability as zero.

We can now define the failure probability vector $x$. To avoid division by zero in cases where node in-degree is zero, we  first define a vector $o$ such that

\begin{equation} \label{eq6.1}
o_{i} = Max\left( \frac{d_{w}^{i}  - k}{k},0 \right),\ \ i \in 1,2,\ldots,N
\end{equation} so that $o_i$ will have a value of $0$ for those nodes which have zero in-degree.  Now for each node, the number of neighbours which failed in time step
\(t - 1\) is given by \(c_{t - 1}A\). The vector of  failure probabilities $x_i$ for all nodes, based
only on neighbour failures in the previous time step, is therefore given
by:

\begin{equation} \label{eq7}
x = c_{t - 1}A\ \bigodot\left( \frac{1}{o_{1} + 1},\frac{1}{o_{2} + 1},\ldots\frac{1}{o_{N} + 1} \right)
\end{equation} where \(\bigodot\) is the Hadamard product. Note that  for any node with a non-zero in-degree, the value of $\frac{1}{o_{i} + 1}$ will be $\frac{k}{d_{w}^{i} }$,  as dictated by  the proportionality properties mentioned earlier.

This does not take into account the effect of neighbours which failed in
time steps prior to \(t - 1\). The effect of these previous failures is
calculated using the probability vector from the previous time step,
\(p_{t - 1}\). For each element of \(p_{t - 1}\), this probability is
first discounted by \emph{r} , given by \(\frac{p_{t - 1,i}}{1 + r}\).

In this model, independence is assumed between the impact of failed
neighbours. Therefore, the total probability of failure for each element
\(p_{t,i}\), can be calculated by:

\begin{equation} \label{eq8}
p_{t,i} = \frac{{x_{i}p}_{t - 1,i}}{1 + r} + x_{i}\left( 1 - \frac{p_{t - 1,i}}{1 + r} \right) + \frac{(1 - x_{i})p_{t - 1,i}}{1 + r}
\end{equation}

As such, the probability vector at time \emph{t} is given by:

\begin{equation} \label{eq9}
p_{t} = x + \frac{1}{(1 + r)}\left( p_{t - 1}\bigodot(1 - x) \right)
\end{equation}

Once the probability vector has been computed, the actual node failures
in time step \emph{t} can be determined stochastically by generating a
vector of random values \emph{z} of length \emph{N}, where
\(0 \leq z_{i} \leq 1,\ i \in 1,2,\ldots,N\). This random vector is then
compared with the probability vector \(p_{t}\), and the cascade vector
\(c_{t}\) is updated according to the result of the comparison:

\[c_{t,i} = \left\{ \begin{matrix}
1 & p_{t,i} < z_{i} \\
0 & \text{otherwise} \\
\end{matrix} \right.\ ,\ \ i \in 1,2,\ldots,N\]

The outcome of this cascading failure simulation is then recorded by updating the failure
matrix \emph{F} with the contents of the cascade vector c, using a
binary OR ($\vee $) operator to carry forward any failures from previous time
steps so that:

\begin{equation} \label{eq12}
f_{t,i} = f_{t - 1,i} \vee c_{t,i},\ \ i \in 1,2,\ldots,N
\end{equation}

\subsubsection{Simulation}

The cascading failure simulation is undertaken on the shareholding network, and initialised with an initial shock equal to 10\% of the number of nodes failing. Mathematically, 0.1 $\times$ N elements of the initial cascade vector $c_0$ are selected at random and set to a value of 1. The remaining elements retain their value of 0. Two types of experiments are then carried out using the cascading failure model described above. The first set of simulations involve  an exploration of the parameter space of $\alpha$ and $\gamma$ by varying these parameters systematically. We did not try to calibrate specific values for these parameters to reflect potential real-world values. Instead, by exploring the parameter space, qualitative assessments are  made about the impact of these parameters on the cascading failure outcomes.

The results of these cascading failure simulations are measured in two ways. The first is by measuring the mean downtime, being the average proportion of the simulation that the average node is down, excluding nodes removed in the initial shock. It is defined and calculated as the sum all of the elements of the failure matrix:

\begin{equation} \label{eq13}
\bar{\tau}=\frac{\sum_{i,j}^{}F_{\text{ij}}}{\text{NT}} - 0.1
\end{equation}

The second measurement is failure proportion, being the total proportion of nodes which failed by the end of the simulation, excluding nodes removed in the initial shock. It is calculated as the sum of the final row of the failure matrix:

\begin{equation} \label{eq14}
\phi=\frac{\sum_{j}^{}F_{T,j}}{N} - 0.1
\end{equation}

In both equations \ref{eq13} and \ref{eq14}, the $0.1$ term obviously represents the initial shock.

The second set of simulations involved an analysis of the country overlap networks of the top 20 countries in the world (by GDP), and whether the overlap network topology has any relationship to the cascading failure outcomes on a country-by-country basis. The countries considered are United States (US), China (CN), Japan (JP), Germany (DE), Great Britain (GB), India (IN), France (FR), Italy (IT), Canada (CA), South Korea (SK), Russia (RU), Brazil (BR), Australia (AU), Spain (ES), Mexico (MX), Indonesia (ID), Netherlands (NL), South Africa (SA), Turkey (TR), and Switzerland(CH).

\section{Results} \label{results}

In this section we present the results of the simulation experiments conducted above.

\subsection{Information theoretic measures}

First we compute information theoretic measures, with the intention of understanding causality of link formation. In other words, the intention is to understand whether the existence of a link between two particular nodes in the co-patenting network makes it more likely for a link to be formed between the same two nodes (firms) in the shareholding network in the following years, and conversely,  whether the existence of a link between two particular nodes in the shareholding network makes it more likely for a link to be formed between the same two nodes (firms) in the co-patenting network in the following years. Therefore, to compute these information  theoretic measures, we never directly compare the co-patenting network and the shareholding network from the exact same year. Rather, we compare the co-patenting network of one year (say, 2008), with shareholding networks from several following years (2009, 2010... up to 2016), and conversely,  we compare the shareholding network of one year (say, 2008), with co-patenting networks from several following years (2009, 2010... up to 2016).

\subsubsection{Mutual Information measures}

\begin{figure*}[htbp]
\centering
		\includegraphics[scale = 0.4]{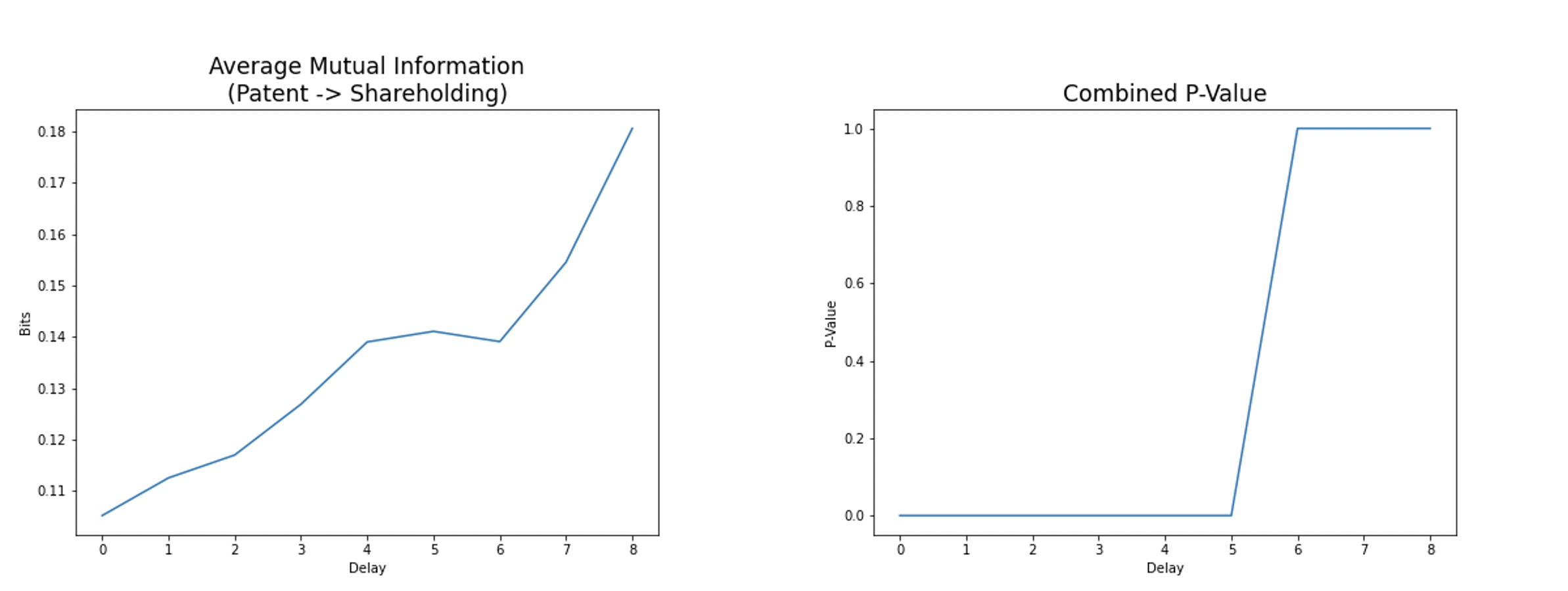}
\caption{The mutual information between edge formation in the co-patenting network and  shareholding networks from later years.  The time difference (delay) between the networks is mentioned in years, indicating that the shareholding network is from the later year, by that number of years.} 
\label{fig5A}
\end{figure*}

\begin{figure*}[htbp]
\centering
		\includegraphics[scale = 0.4]{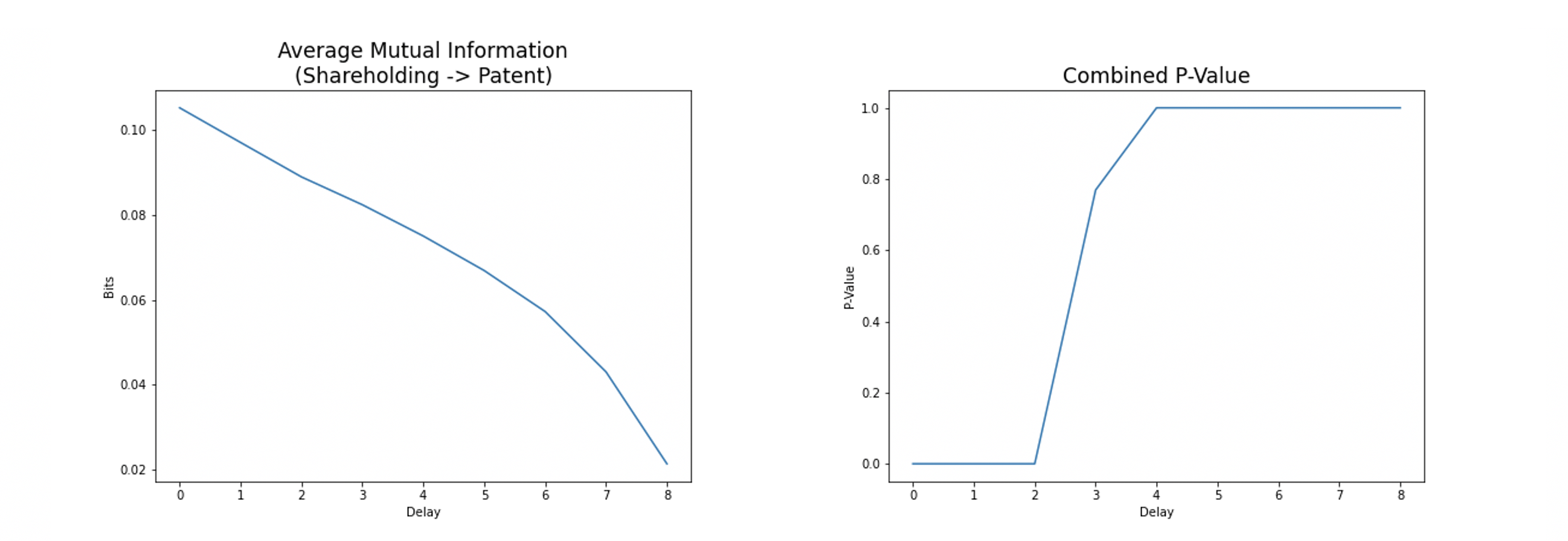}
\caption{The mutual information between edge formation in the shareholding network and  co-patenting networks from later years. The time difference (delay) between the networks is mentioned in years, indicating that the co-patenting network is from the later year, by that number of years. } 
\label{fig6A}
\end{figure*}

Fig.  \ref{fig5A} shows the mutual information between edge formation in the co-patenting network and  shareholding networks from later years. It can be seen that the amount of mutual information tends to increase with greater amounts of delay (difference in years), and the p-value is significant upto a difference of five years. On the other hand,  Fig.  \ref{fig6A} shows the mutual information between edge formation in the shareholding network and  co-patenting networks from later years. Since the mutual information is a symmetric measure, the same result of approximately 0.105 bits is found with a delay of 0 -  that is, when both networks are from the same year. However, the mutual information then tends to decrease with greater amounts of  year difference. The p-value is significant when the year difference is upto  two years.

These results suggest that  a patent link between two nodes in the co-patenting network may cause a shareholding link between the same two nodes (firms) in the coming years, and not vice versa. In other words, the directionality of the link creation mechanism is from patent link to shareholding link, rather than shareholding link  to patent link. The result is statistically significant up to a delay of 5 years in the case of the relationship in Figure 5. However, this is may be due to limitations in the available data. Since only 9 years of data are available, with greater amounts of delay, there become fewer data points. (For example, with a delay of 5 years, there are only 4 data points available for each edge.) Therefore, the p-values only indicate here the limitation in our dataset, and in reality, the causal relationship may well exist beyond the five years suggested by these results.

\subsubsection{Intra-national vs international link formation and mutual information}

\begin{figure*}[htbp]
\centering
		\includegraphics[scale = 0.4]{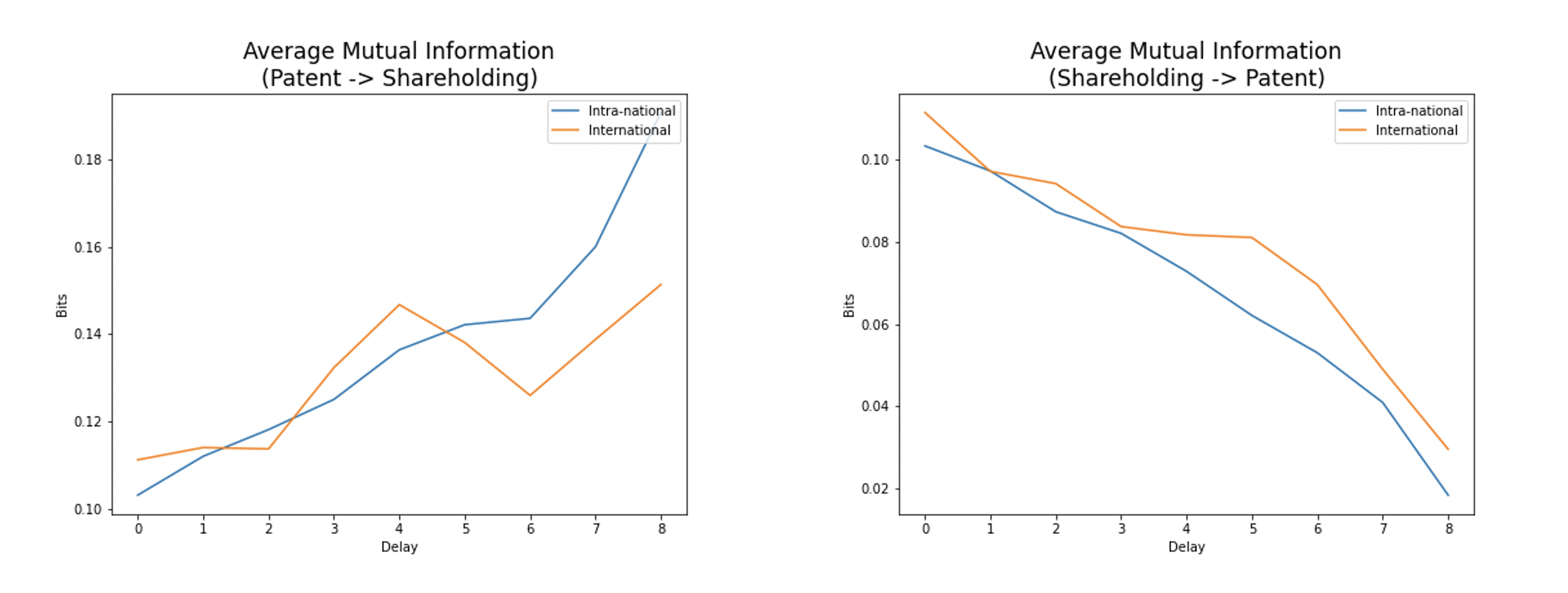}
\caption{The  mutual information between the the co-patenting networks and later (in terms of year considered) shareholding networks,  and vice-versa, for both intra-national and international links} 
\label{fig11A}
\end{figure*}

Since the mutual information results suggested a directionality to the relationship from patent data to shareholding data, further analysis was then carried out to determine whether this varied based on  whether the relationship was between two firms from the same country, or two firms from different countries. Fig.  \ref{fig11A} plots the mutual information between the co-patenting networks and later (in terms of year considered) shareholding networks, for both intra-national and international links, as well as the mutual information between the shareholding networks and later (in terms of year considered) patent  networks, for both intra-national and international links.

From  Fig.  \ref{fig11A}, It can be seen that the  trend seen in Fig.  \ref{fig5A} was actually disguising two separate trends. For international links, there is a peak in mutual information with a year-difference of 4 years, whereas the mutual information for intra-national links continues to increase as the year-difference increases. This indicates that there may be a longer delay, in terms of lapsed years, between the formation of patent collaboration relationships and shareholding relationships at the domestic level than at the international level. The reason for this is not clear, but it is likely that firms with international relationships would have different characteristics  and attributes from those which are more domestically focussed. For example, firms with international links may be larger with greater financial resources, and therefore more willing and able to make investments in other firms over shorter time periods.

\subsubsection{Country analysis based on mutual information}

\begin{figure*}[htbp]
\centering
		\includegraphics[scale = 0.5]{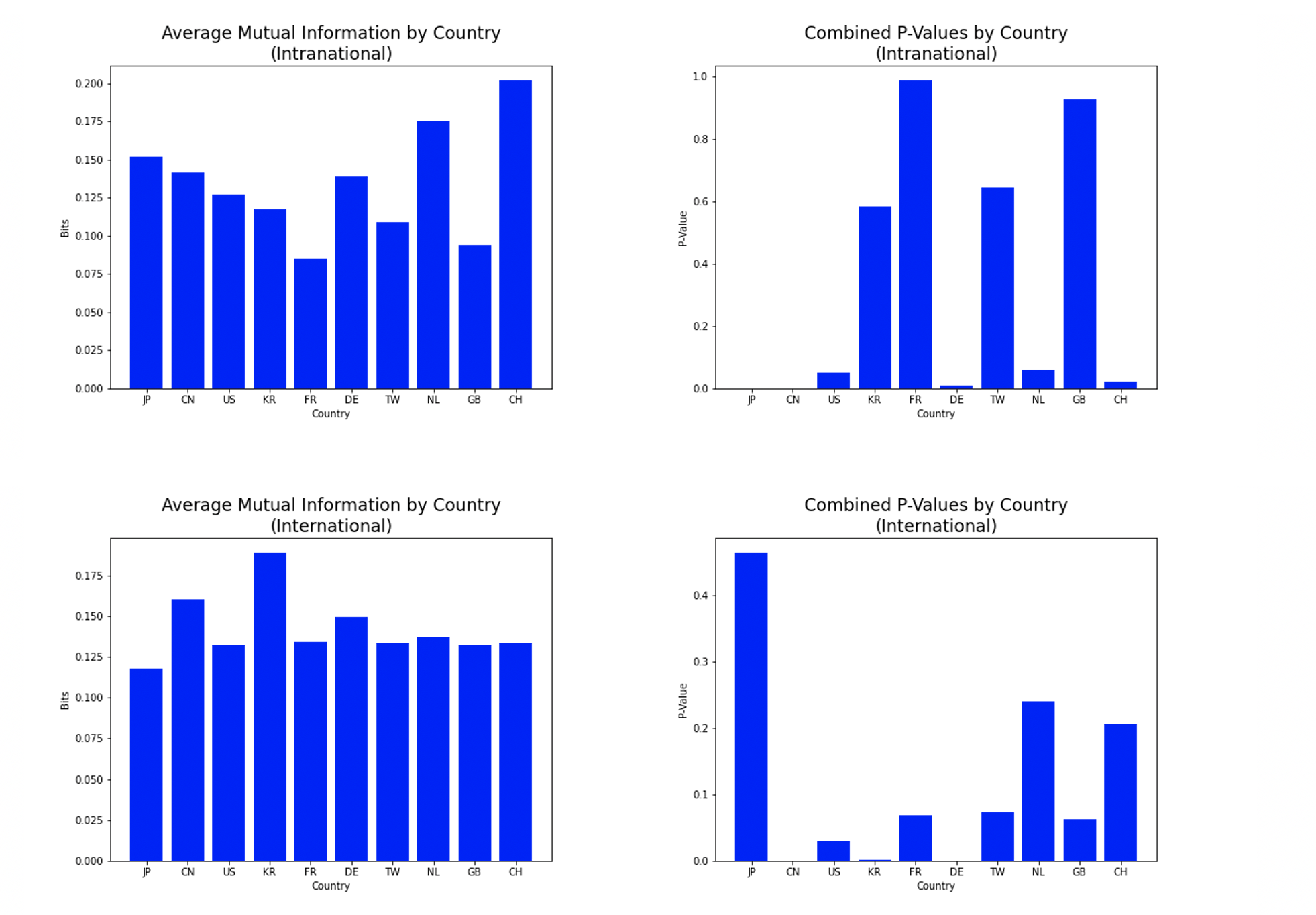}
\caption{The  mutual information values for intra-national links and international links for  ten selected countries or autonomous regions:  Japan (JP), China (CN), United States (US), South Korea (KR), France (FR), Germany (DE), Taiwan (TW), Netherlands (NL), Great Britain (GB), and Switzerland (CH). } 
\label{fig12A}
\end{figure*}

Finally, we measured mutual information between shareholding networks and co-patenting networks on a country-by-country basis for the 10 countries or autonomous regions appearing most frequently in the co-patenting networks: Japan (JP), China (CN), United States (US), Korea (KR), France (FR), Germany (DE), Taiwan (TW), Netherlands (NL), Great Britain (GB), and Switzerland (CH).  Fig.  \ref{fig12A} shows these measurements. The country-by-country analysis yielded statistically significant results for some countries and not others, as the corresponding p-values indicate. For those countries with statistically significant results ($p \le 0.5$), there is a range of mutual information values. For example, Switzerland has a much higher mutual information result for intra-national links than Japan, China, the US or Germany (which all have similar measurements). This may suggest that there are unique factors in relation to research collaboration relationships in Switzerland, which lead to such relationships being more likely to develop into shareholding relationships. On the other hand, some countries which had statistically significant results for intra-national links did not have a statistically significant result for international links. For example, the mutual information  values for Japan indicate that there is a relationship between patent collaboration and shareholdings at the intra-national level, but not at the international level. This could be due to particular cultural factors within Japan, such as features of the domestic Keiretsu system~\cite{mcguire2009japanese}. However, in the case of Korea the values (which might similarly be expected to be influenced by the domestic Chaebol system~\cite{campbell2002corporate}) are the opposite, indicating a relationship between patent collaboration and shareholdings at the international but not the intra-national level. This could perhaps be due to differences in the shareholding network topology created by those different systems. Further analysis, perhaps bolstered by additional data, is needed to develop deeper insights in this regard.

\subsubsection{Active Information Storage measures}

\begin{figure*}[htbp]
\centering
		\includegraphics[scale = 0.4]{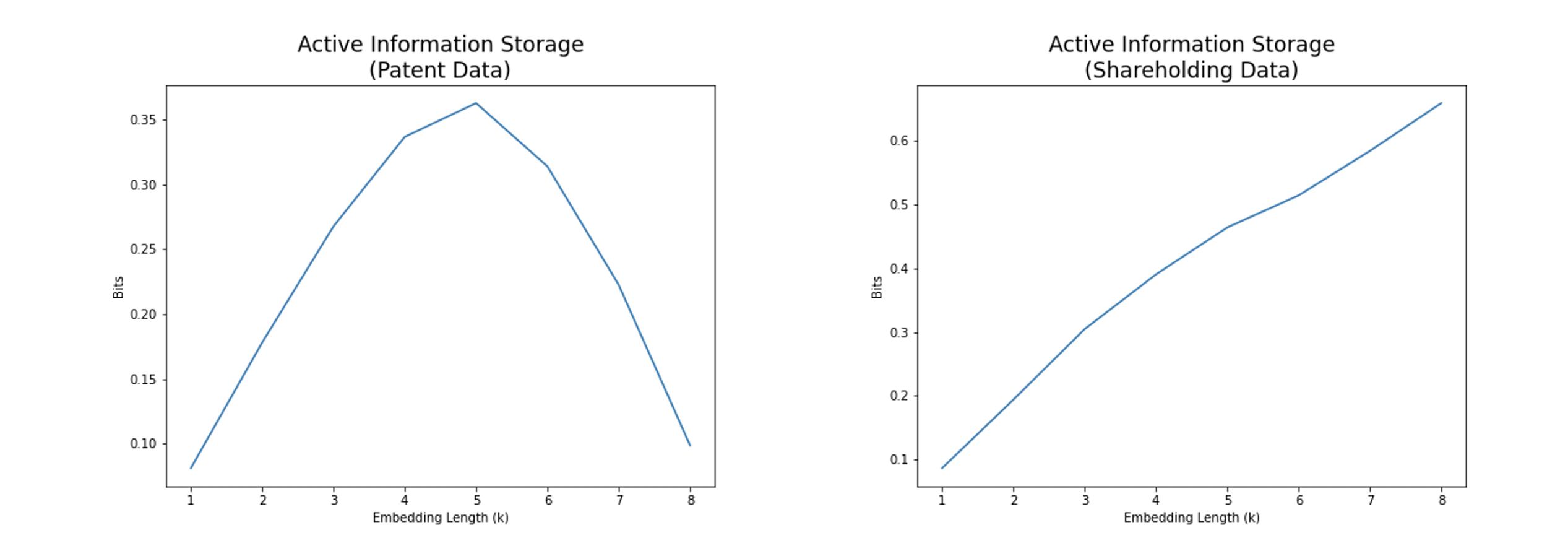}
\caption{The active information storage for the patent and shareholding networks from different years.} 
\label{fig7A}
\end{figure*}

Next we computed the transfer entropy  between the co-patenting networks and the shareholding networks from various years (ranging from 2008 to 2016). However, in order to determine the appropriate target embedding length for transfer entropy calculations, the active information storage was first calculated for each variable with various amounts of delay (year difference). Fig.  \ref{fig7A} shows the active information storage for the patent and shareholding networks from different years.

It can be seen that, for the co-patenting networks, there is a peak at k=5, indicating a five year interval between co-patenting network topologies result in peak active information storage in terms of link structure. An embedding length of 5 is therefore used for the following transfer entropy calculations using the patent data as the target variable.

However, the figure also  shows that for the shareholding networks across different years, there is no peak in active information storage, and the active information storage continues to increase up to a difference of 8 years, the maximal year difference possible between datasets corresponding to all years from 2008 to 2016. Longer time series data would likely be required in order to determine the year difference for which active information storage is maximised in terms of link structure of shareholding networks.  Since we do not have such data, for the purposes of this work, a difference of 8 years is used in our transfer entropy calculations as described below, which maximises active information storage based on the available data.

In addition to determining the target embedding length for transfer entropy calculations, the active information storage results also provide insight into some other aspects of the interplay between the evolutions of interfirm co-patenting and shareholding relationships.  Namely, these results suggest that the most information about current patent collaboration relationships is gained by looking back 5 years. However, the most information about current shareholding relationships can be gained by looking back much further (at least 8 years, and possibly more).  This suggests that interfirm shareholding relationships may  evolve relatively slowly, whereas patent collaboration relationships impact  the future evolution of corresponding network topology in  typically shorter term, perhaps in the vicinity of 5 years.

\subsubsection{Transfer Entropy measures}

\begin{figure*}[htbp]
\centering
		\includegraphics[scale = 0.4]{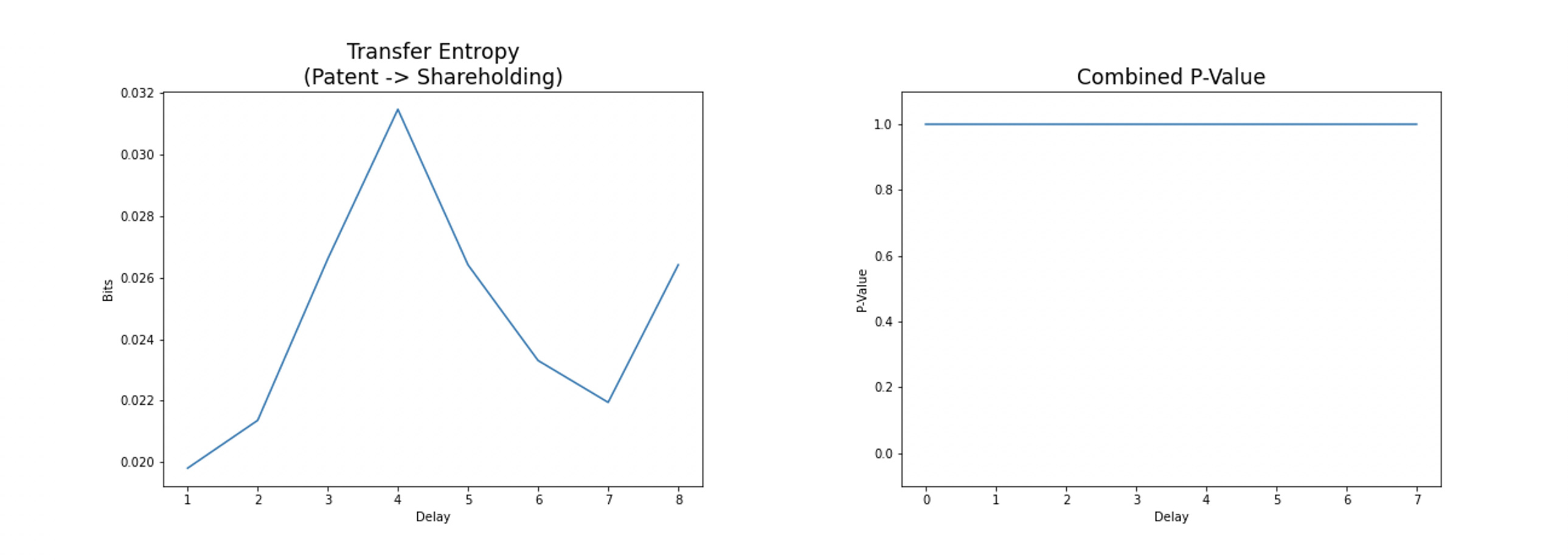}
\caption{The transfer entropy between the co-patenting network and later  shareholding networks using a target embedding length of 8.  In other words, the calculation of transfer entropy between co-patenting links and shareholding links which appear eight years later.} 
\label{fig8A}
\end{figure*}

\begin{figure*}[htbp]
\centering
		\includegraphics[scale = 0.4]{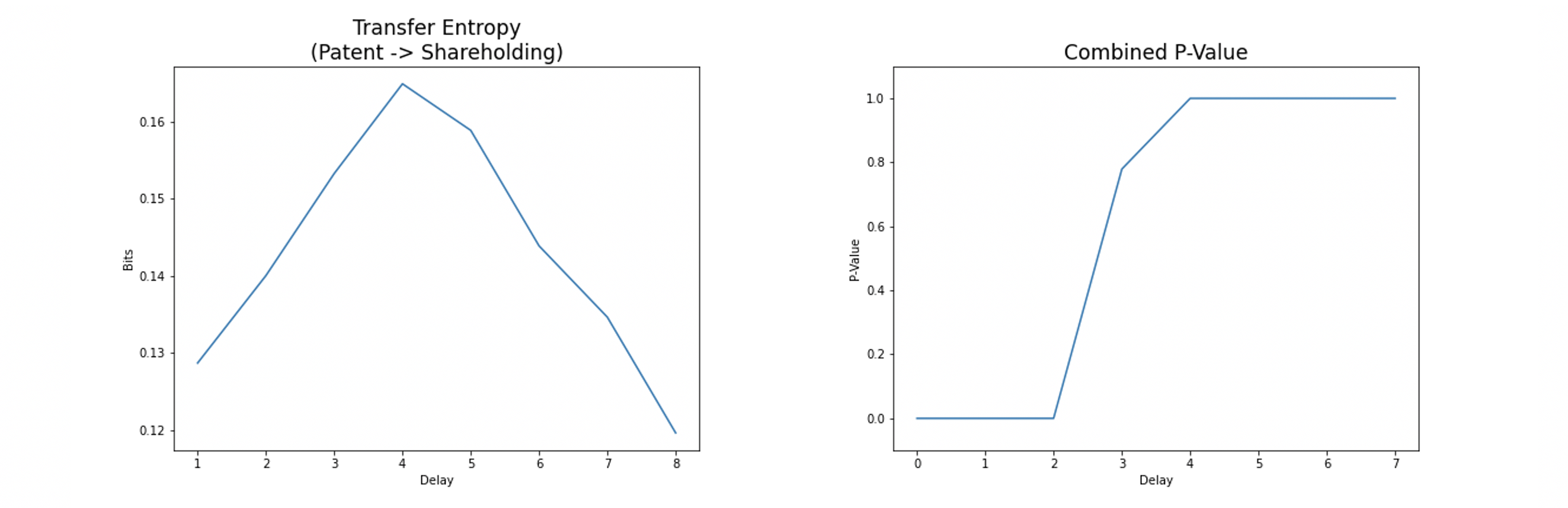}
\caption{The transfer entropy between the co-patenting network and later  shareholding networks using a target embedding length of 1.  In other words, the calculation of transfer entropy between co-patenting links and shareholding links which appear one year later.} 
\label{fig9A}
\end{figure*}

%Having established the optimal target embedding length, 

We made  transfer entropy calculations for various delay lengths between patent and shareholding networks (and vice versa). Fig.  \ref{fig8A} shows the transfer entropy results between the co-patenting networks and shareholding networks using a target embedding length of 8 (the links considered from the shareholding network are eight years later than the links considered in co-patenting network). The results seemingly indicate a peak in transfer entropy with a source-target difference of 4 years. However, with only nine years of data available, these results were not statistically significant. Meanwhile, Fig.  \ref{fig9A} shows the transfer entropy results between the co-patenting network and shareholding networks with a target embedding length of 1 (the links considered from the shareholding network are only one year later than the links considered in co-patenting network). Although  it should be noted that these calculations are not conditional upon the history of the shareholding data which maximises the amount of stored information in that variable, the results do show a similar peak with source-target delay of 4 years. In this case, statistically significant results were obtained up to a delay of 2 years.

It is possible that the lack of statistically significant results for some transfer entropy calculations is due to limitations in the number of years for which data is available.  It may be that with a longer time series, more robust conclusions could be drawn in relation to the transfer entropy. Nevertheless, it is significant that there is a significant amount of transfer entropy between shareholding network data (as target variable) and co-patenting network data (as source variable), indicating that there is information transfer from co-patenting network link formation to shareholding network link formation, which peaks when the time difference between these two processes is about four years.

\subsection{Cascading failure analysis of the shareholding networks}

Here we present the results of the cascading failure simulations, as described in section \ref{methodology}, undertaken on the shareholding networks. Recall that we defined two key parameters $\alpha$ and $\gamma$ in section  \ref{methodology},  which represent the cumulative failure rate and the overall discount rate of failure probability in terms of cascading failures for shareholding networks. 

\subsubsection{Mean downtime and mean node failure proportion during cascading failure of shareholding networks}

\begin{figure}[htbp]
\centering
		\includegraphics[scale = 0.3]{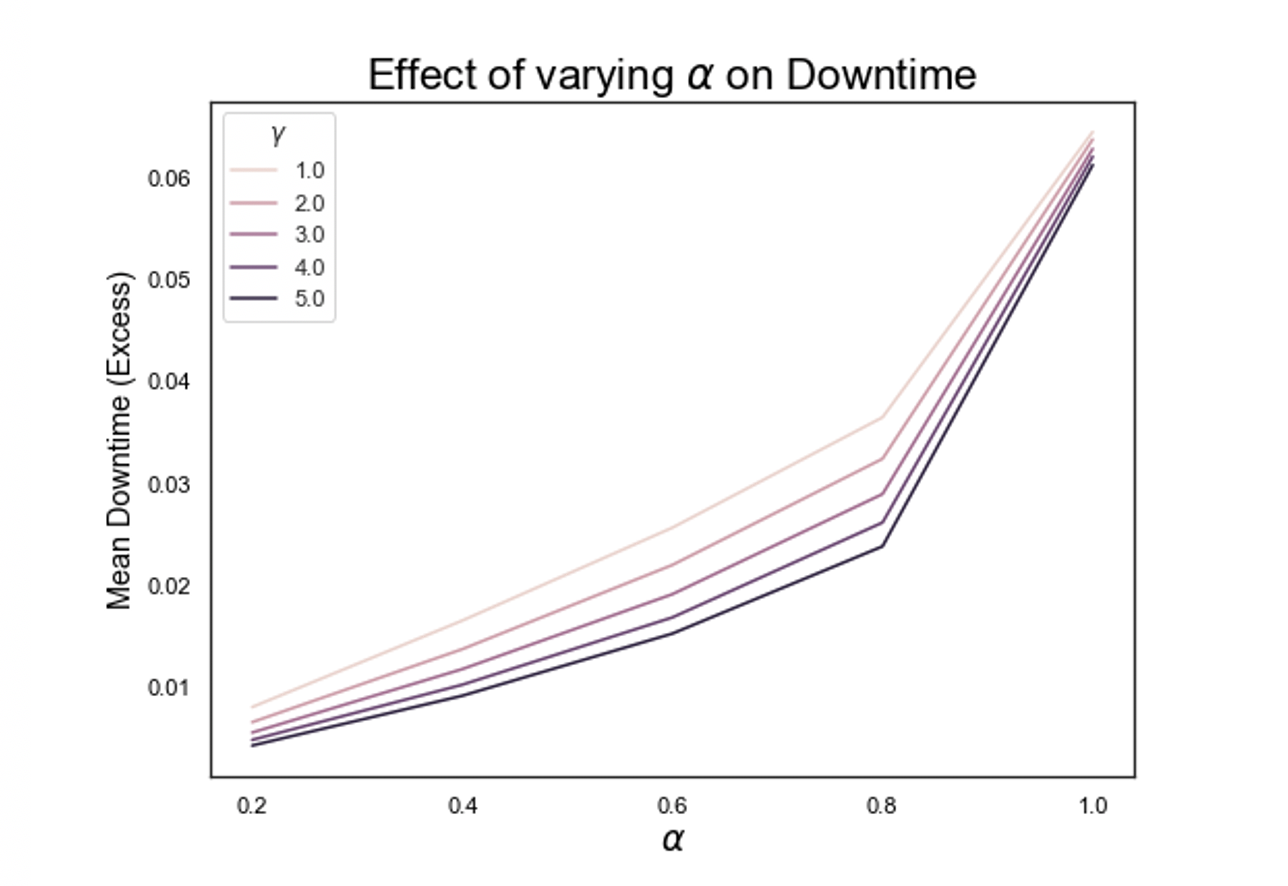}
\caption{Mean downtime vs  cumulative failure rate $\alpha$ during cascading failure simulation of shareholding networks. The mean is obtained by averaging across all years for which data is available (2008 -  2016).} 
\label{fig2B}
\end{figure}

\begin{figure}[htbp]
\centering
		\includegraphics[scale = 0.3]{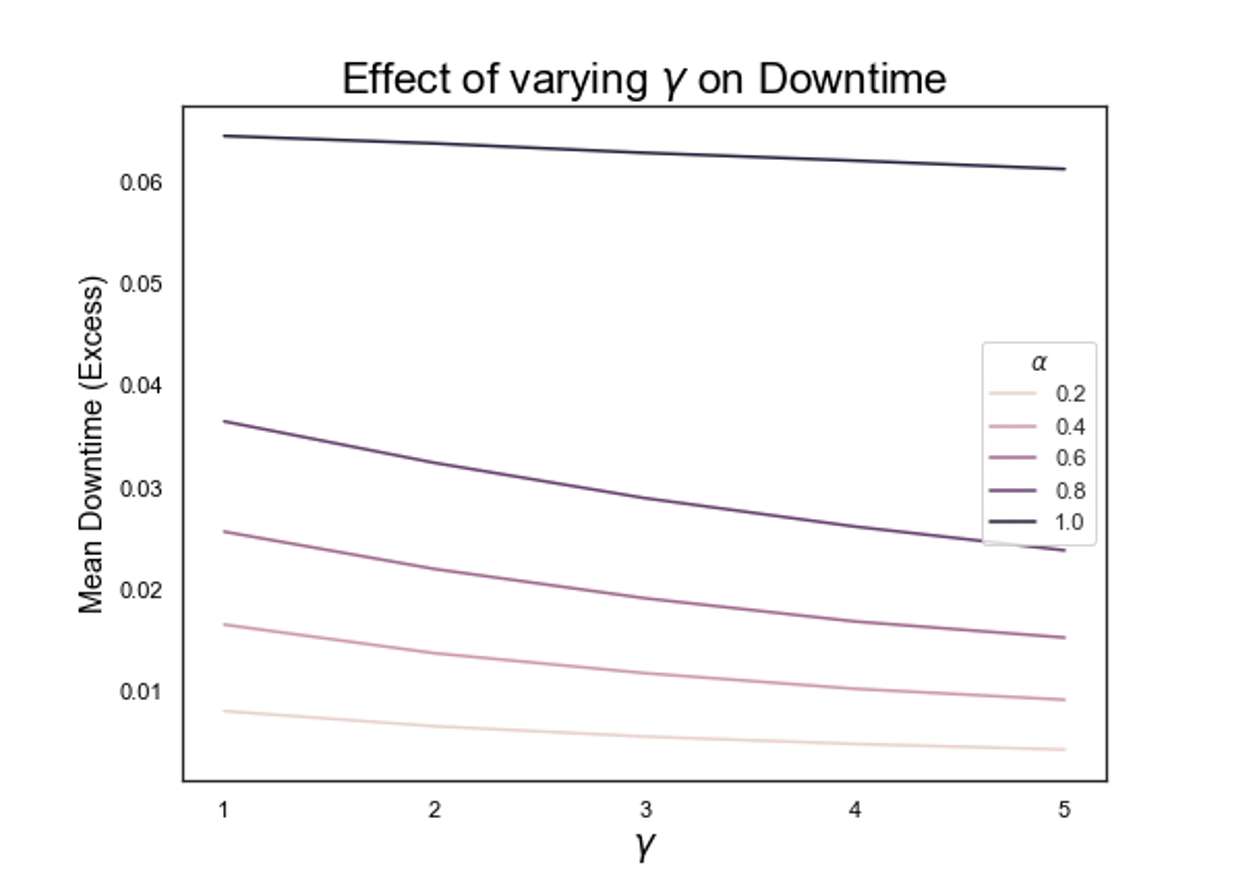}
\caption{Mean downtime vs  the overall discount rate $\gamma$ during cascading failure simulation of shareholding networks. The mean is obtained by averaging across all years for which data is available (2008 -  2016).} 
\label{fig3B}
\end{figure}

\begin{figure}[htbp]
\centering
		\includegraphics[scale = 0.3]{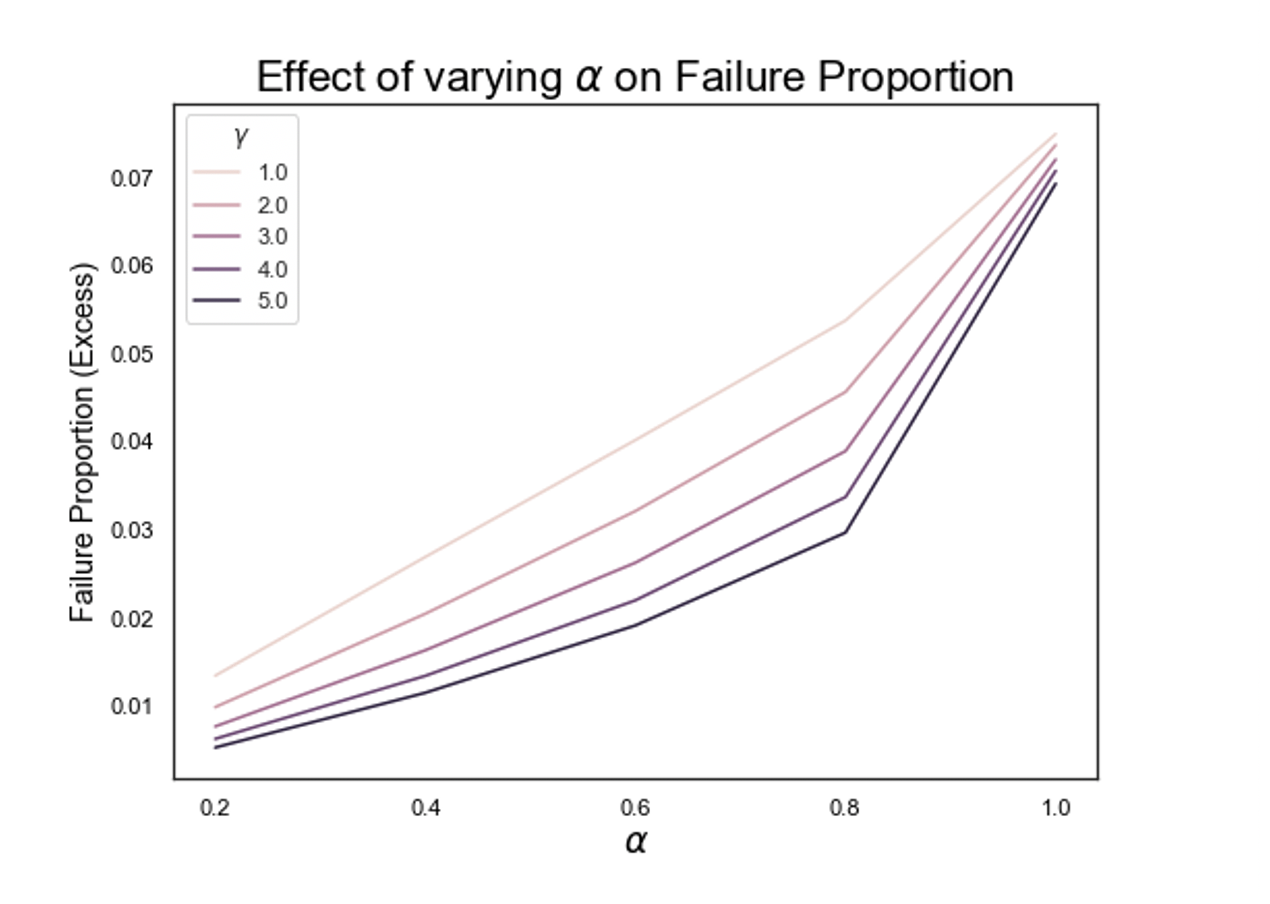}
\caption{The failure proportion  vs  cumulative failure rate $\alpha$ during cascading failure simulation of shareholding networks. The average is obtained by averaging across all years for which data is available (2008 -  2016).} 
\label{fig4B}
\end{figure}

\begin{figure}[htbp]
\centering
		\includegraphics[scale = 0.3]{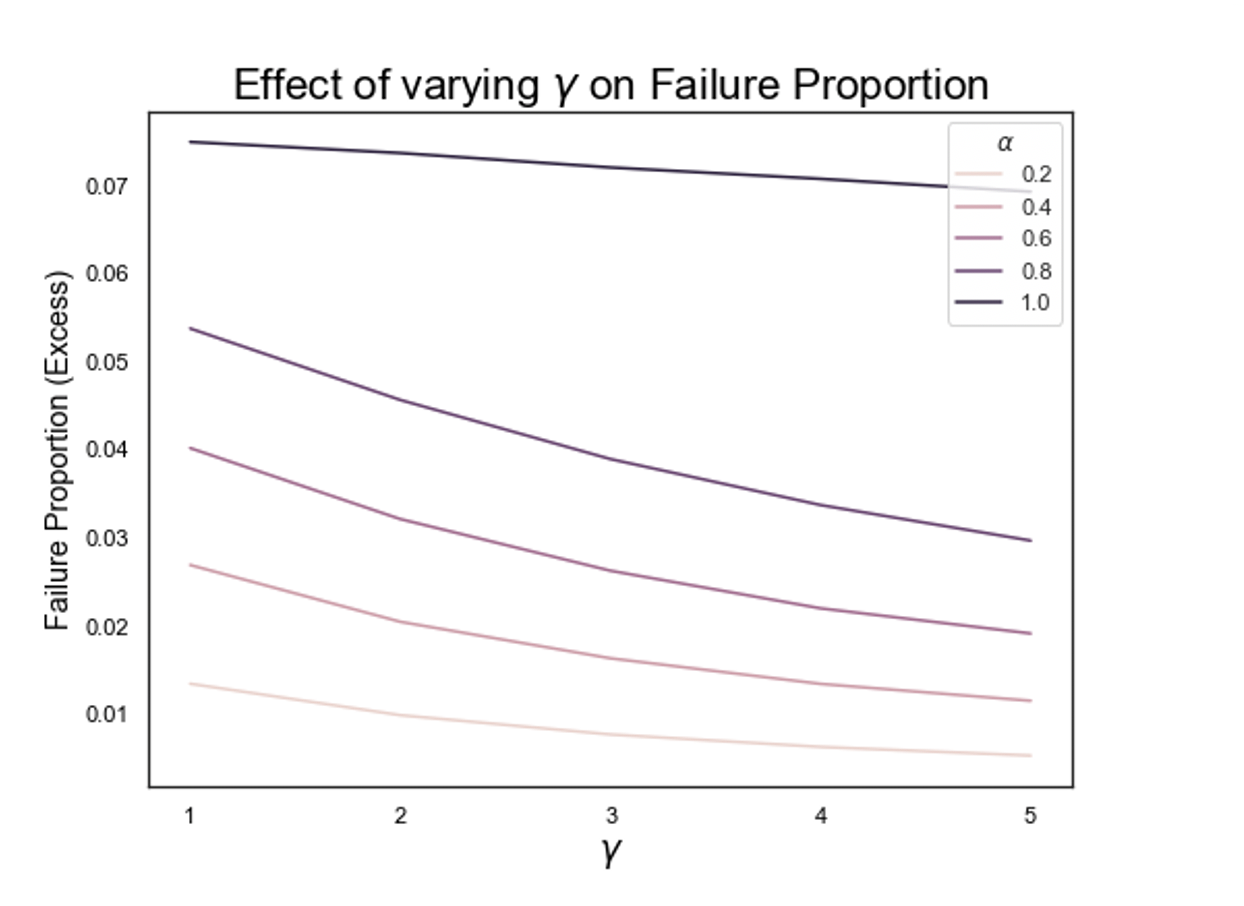}
\caption{The   failure proportion vs  the overall discount rate $\gamma$ during cascading failure simulation of shareholding networks. The mean is obtained by averaging across all years for which data is available (2008 -  2016).} 
\label{fig5B}
\end{figure}

We consider the mean downtime, and the node failure proportion, of shareholding networks, as the parameters  $\alpha$ and $\gamma$  (the cumulative failure rate and the overall discount rate) are varied. We varied   $\alpha$ from $0.2$ to $1.0$ with a step increase of $0.2$, and $\gamma$ from 1 to 5 with a step increase of 1. For each combination of  $\alpha$ and $\gamma$, we conducted the cascading failure simulation as described in section \ref{methodology} for all shareholding networks, and calculated the mean down time and the node failure proportion (across networks from all considered years: i.e, from 2008 to 2016), as given by equations \ref{eq13} and \ref{eq14} respectively.  The results of these experiments are shown in figures \ref{fig2B},  \ref{fig3B},  \ref{fig4B},  \ref{fig5B}.

\begin{figure*}[htbp]
\centering
		\includegraphics[scale = 0.3]{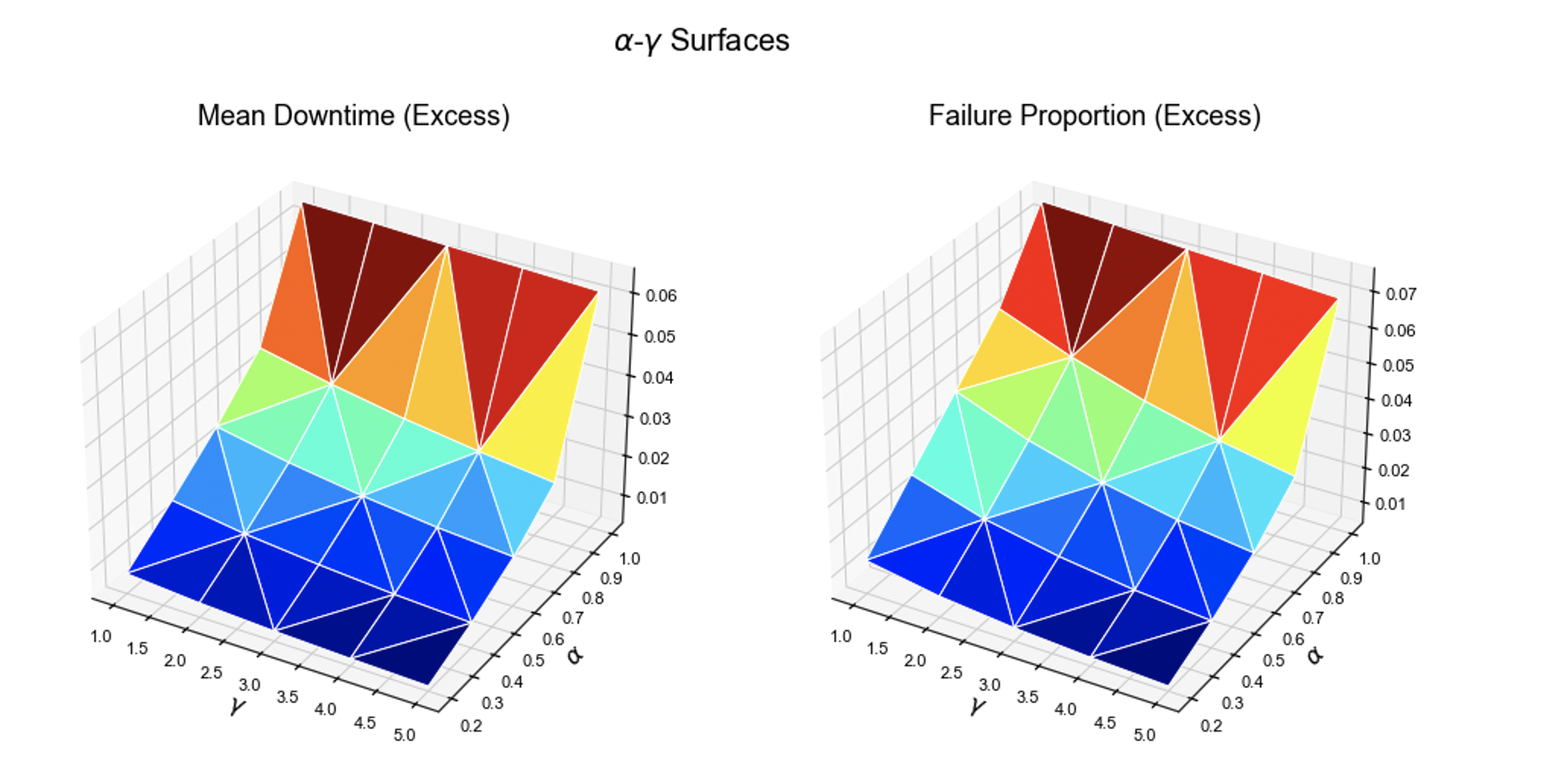}
\caption{The  mean downtime and the  failure proportion against cumulative failure rate $\alpha$  and overall discount rate $\gamma$ shown as surface plots. } 
\label{fig6B}
\end{figure*}

According to figures \ref{fig2B} -- \ref{fig5B}, the mean downtime increases when the cumulative failure rate increases. This is not surprising. What is interesting is that when this cumulative failure rate exceeds $0.8$, there is a sharp rise in mean downtime. Similarly,  when the overall discount rate increases, the mean downtime decreases. Again, this is not surprising given that the higher the discount rate, the lower the failure probabilities become over the years, even if they start high to begin with.  Very similar trends are observed in terms of node failure proportions. These results can also be visualised as a surface plot of the parameters $\alpha$ and $\gamma$, as shown in Fig.  \ref{fig6B}.

\subsubsection{Country-based variation of mean downtime and the node failure proportion during cascading failures}

%\begin{figure}[htbp]
%\centering
%		\includegraphics[scale = 0.3]{cutfigures/fig7B.png}
%\caption{Time series plots of the network size for each country subnetwork} 
%\label{fig7B}
%\end{figure}
%
%\begin{figure}[htbp]
%\centering
%		\includegraphics[scale = 0.3]{cutfigures/fig8B.png}
%\caption{Time series plots of the network size for each country subnetwork} 
%\label{fig8B}
%\end{figure}

\begin{figure}[htbp]
\centering
		\includegraphics[scale = 0.3]{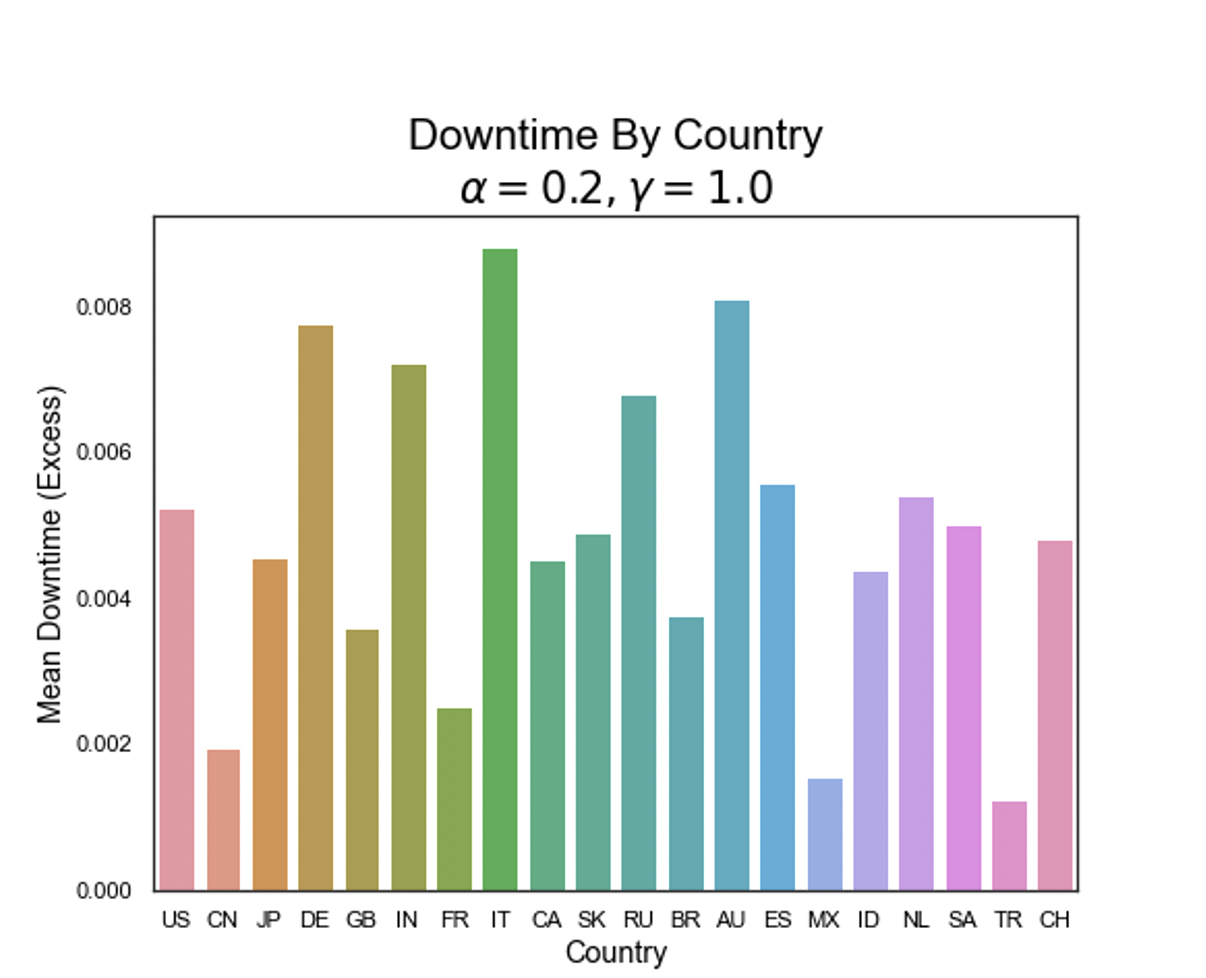}
\caption{Mean downtime based on country-specific shareholding networks during cascading failures. Cumulative failure rate $\alpha=0.2$  and overall discount rate $\gamma=1.0$  is used. } 
\label{fig11B}
\end{figure}

\begin{figure}[htbp]
\centering
		\includegraphics[scale = 0.3]{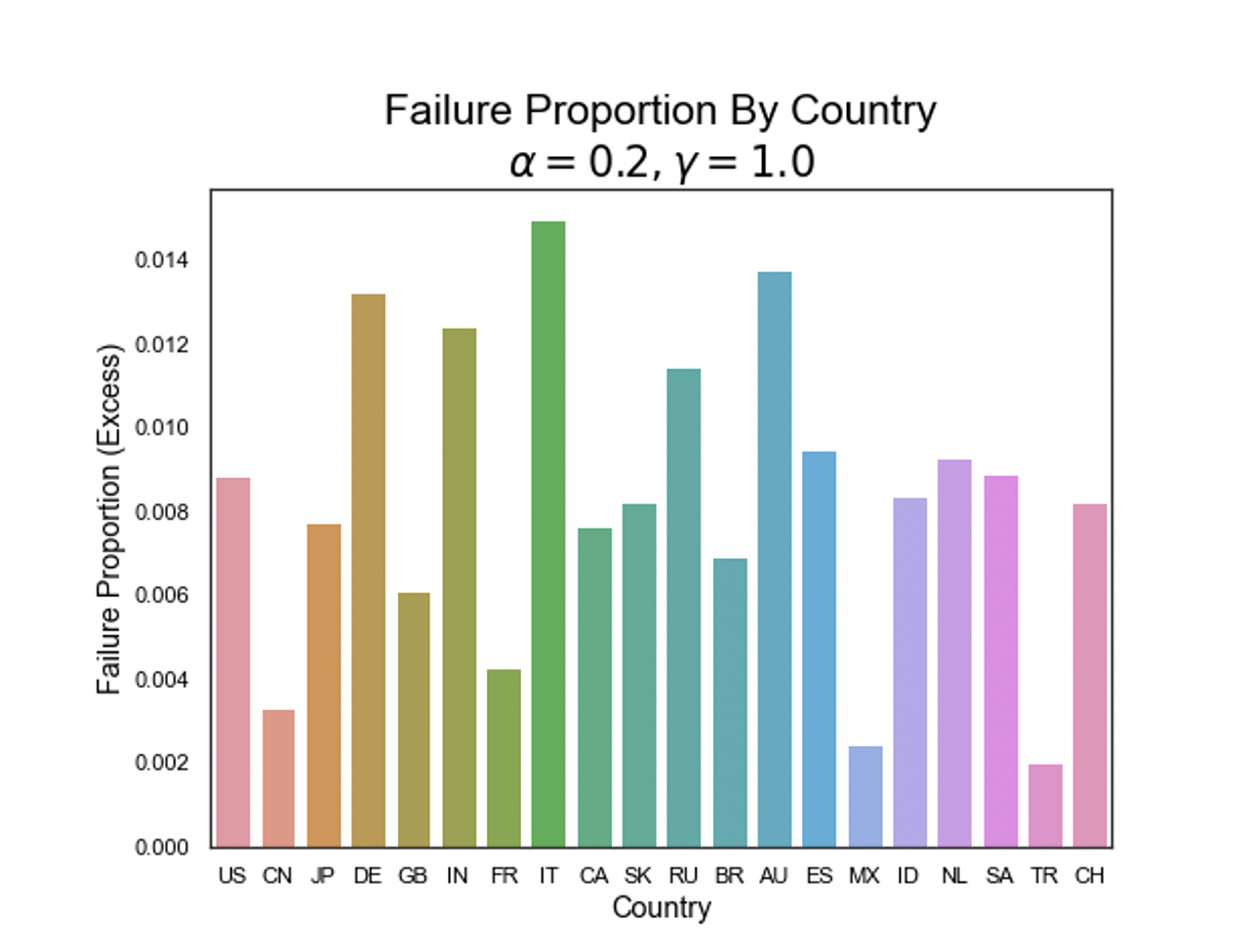}
\caption{The node failure proportion based on country-specific shareholding networks during cascading failures. Cumulative failure rate $\alpha=0.2$  and overall discount rate $\gamma=1.0$  is used.} 
\label{fig12B}
\end{figure}

It is also interesting to note that mean downtime and the failure proportion can vary a lot based on countries. Therefore, we constructed country-specific subnetworks of shareholding networks, and simulated cascading failures on them, again for a range of   cumulative failure rate $\alpha$  and overall discount rate $\gamma$ values. Of course, the values  of mean downtime and node failure proportion vary depending on the values  $\alpha$  and  $\gamma$  as described already, but we also observed country-specific variations. For example, we show the results for  $\alpha=0.2$  and  $\gamma=1.0$  (the lowest values considered for these parameters) for a number of countries in  Fig. \ref{fig11B}  and Fig. \ref{fig12B}.  As mentioned above, these countries are selected on the basis that these are the top twenty economies in the world by GDP. Based on these figures, we could observe that certain country-specific shareholding networks, such as those of Italy and Australia, have relatively high mean downtime, while  other  country-specific shareholding networks, such as those of Mexico and Turkey, have really low downtime. Similarly, in terms of node failure proportion, some countries such as Mexico and Turkey again have the lowest node failure proportions, meanwhile countries such as Australia and Italy, have the highest node failure proportion. It is interesting to know that these two metrics of resilience against cascading failures that we use give more or less identical results  in terms of the relative order of countries in terms of their resilience. Therefore it could be concluded that some country-specific shareholding networks are a lot more vulnerable to cascading failures compared to others, and it is concerning to note that certain economic superpowers, such as Germany, India, and the US, have relatively less resilient country-based shareholding networks in terms of the above-mentioned metics. Further analysis on this aspect, possibly with further data, is necessary to reach more definite conclusions in this regard, and this is beyond the scope of this paper.

It is important to recall here that in constructing the shareholding networks, the shareholding links were not weighted: that is, the value of the shares was not considered. If a firm $A$ owns shares in firm $B$, then it was assumed that a directed shareholding link exists between $A$ and $B$. It is this network that is used to simulate the cascading failure model. This is an important limitation, as the failure probability of firm $A$ if firm $B$ has failed would very much depend on the value of the shares firm $A$ holds in firm $B$.  Nevertheless, the results we present here are system-level results, rather than results pertaining to individual firms. Namely, our focus is on downtime and failure proportion, which are both properties of the shareholding network as a whole, or country-specific shareholding networks, and not individual firms or small groups of firms. For the very large networks involved in this study, it could be argued with some certainty that the effects of varying shareholding link weights would be averaged out at the network level, and the  downtime and failure proportion parameters measured at network level are not severely affected by the variability in individual link weights which were not considered. Nevertheless, it it acknowledged that, if data is available, a more accurate cascading failure model could be built by considering the link weights which represent the value of the shares in each shareholding relationship.

%Time series plots of the network size for each country subnetwork are shown in Fig. \ref{fig7B} and Fig. \ref{fig8B} . The mean downtime and failure proportion results are then shown on a country-by-country basis in Fig. \ref{fig11B}  and Fig. \ref{fig12B} .
%
%Topological features (average in-degree and estimated power law exponent) of each country subnetwork are then shown in  Fig. \ref{fig13B} and   Fig. \ref{fig14B}. The results of linear regression analyses using these features are shown in Tables 2 and 3, with regression plots shown in Figures \ref{fig15} and \ref{fig16} respectively.

%
%
%\begin{figure}[htbp]
%\centering
%		\includegraphics[scale = 0.3]{cutfigures/fig13B.png}
%\caption{Average in-degree by country} 
%\label{fig13B}
%\end{figure}
%
%
%\begin{figure}[htbp]
%\centering
%		\includegraphics[scale = 0.3]{cutfigures/fig14B.png}
%\caption{Power-law exponent by country} 
%\label{fig14B}
%\end{figure}
%
%
%
%\begin{figure}[htbp]
%\centering
%		\includegraphics[scale = 0.3]{cutfigures/fig15.png}
%\caption{Regression plot for average in-degree} 
%\label{fig15}
%\end{figure}
%
%\begin{figure}[htbp]
%\centering
%		\includegraphics[scale = 0.3]{cutfigures/fig16.png}
%\caption{Regression plot for power-law exponent} 
%\label{fig16}
%\end{figure}

\section{Conclusions} \label{conclusions}

\subsection{Summary}

In this paper, we considered interfirm networks constructed from the Orbis dataset. We constructed shareholding networks and patents networks,   while we also considered the `overlap' networks on a longitudinal basis. We employed information measures  to analyse these networks, with the view of determining if the existence of one type of relationship influences link formation in terms of the other type in these networks. We also simulated cascading failure on shareholding networks, and observed the resilience of these shareholding networks during cascading failures both globally and in terms of individual countries, employing a number of metrics to measure this resilience.

The information measures we used provide a number of interesting insights regarding causality between co-patenting networks and shareholding networks. Firstly, measuring mutual information between co-patenting network and shareholding network topologies suggested that the existence of patent links provides information about the existence of future shareholding links, rather than the other way around. Therefore, by analysing co-patenting networks, we could get insights into the future topology of shareholding networks, though, of course, an abstract measure such as mutual information does not provide much granularity. Nevertheless this observation hints at causality in the co-patenting network to shareholding network direction. We measured transfer entropy, a more direct measure of information transfer, to quantify this further, which also confirmed some information transfer from co-patenting activity to shareholding relationships.  However the transfer entropy analysis was constrained by  insufficient time series data to conduct the analysis with the appropriate target embedding length and source-target delay. Despite this, the observations using information measures are sufficient to suggest that there may be a real benefit to considering interfirm networks more holistically, rather than focussing on one type of interfirm network at a time.

We also measured mutual information  between patent and shareholding networks for each country represented in the Orbis dataset.      Interestingly, there was considerable diversity between countries. This  suggested that in case of some countries, the patent and shareholding network topologies are quite similar, and these are quite different for other countries.       A further observation made was that in the case of some countries, the intra-national topological structure between patent and shareholding networks was quite similar while international links were more dissimilar between patent and shareholding networks, while the opposite was true in the case of other countries. We postulate that cultural factors, such as Keiretsu (Japan) and Chaebol (Korea) might influence this difference. However the main observation that can be made here is that calculating mutual information between two types of links (such as co-patenting links and shareholding links) on interfirm networks is a promising line of enquiry to understand and quantify overall topological similarities between various kinds of interfirm networks which could be built from the same set of firms (nodes). This avenue of inquiry may lead to further insights about interfirm relationships which are not  necessarily available through the study of a single type of  interfirm network alone.

We also simulated cascading failure of shareholding networks using a detailed and nuanced simulation method.  We considered the effects of failure rate and  discount rate  on the overall resilience of shareholding networks against cascading failures, both globally and at individual country level. The goal here was to understand how parameters such as failure rate and discount rate  influence the resilience of shareholding networks, and whether we can make any country-specific observations in this regard by considering firms belonging to each country and shareholding relationships between them as standalone networks, though in truth they are all interconnected. We observed that increasing  the failure rate resulted in increases in both  average downtime and  node failure proportion,  whereas increasing the discount rate resulted in decreases in both  average downtime and  node failure proportion.   This is in accordance with expectations.  A higher probability of failure should lead to more node failures overall. Similarly, a lower discount rate means that the strong likelihood of failure persists further into the future, and so should also lead to more node failures overall. On the whole, the results from  both measurements (average downtime and node failure proportion) appear qualitatively  similar.

Overall, the results of the cascading failure simulations in shareholding networks indicate that if firms are able to recover more quickly from the initial shock (corresponding to a high discount rate-$\gamma$), then the cascade effect will be more limited. For comparison, it could be noted that although the global financial crisis (GFC) did not primarily result from the breaking up of shareholding networks, it is an example of an economic shock which caused  cascading failures, and in which events moved relatively rapidly, (with a lower discount rate) leaving firms with little opportunity to adapt.  In such scenarios, the cascade effect  would be more pronounced, as our simulations with a low value of discount rate indeed indicate. Nevertheless, it should be noted that there were other factors such as government interventions at play during the global financial crisis, which our simple cascading failure model did not capture.

Then we analysed the resilience of individual country-based shareholding  networks  to cascading failures. It was observed that some country networks were quite resilient and others less so to cascading failures - there was a range of values for the downtime and node failure proportion metrics. Nevertheless, it was noted that for any given country, these two measures gave qualitatively similar results, and the order of countries in terms of resilience did not change much based on which resilience metric was used in the analysis. However, the comparison between countries provided important insights for the future stability of global financial markers -  for example, it was observed that according to our simulation experiments, US firms  would experience  much higher downtime and node  failure proportion than Chinese firms during such cascading failure events.

\subsection{Future work}

The presented work had a number of important limitations, primarily due to the limitations in Orbis dataset data that we had access to. This included the limited number of years for which data was available, which in turn affected the information theoretic calculations.  The sheer size of the patent and shareholding complex networks presented another limitation, especially in terms of calculating complex network metrics, so that metrics such as centrality measures could not be considered. In terms of results,  mutual information based measures established that there is shared information between the topologies of co-patenting networks, and shareholding networks from later years, hinting at causality from co-patenting networks to shareholding networks, but information transfer measures such as transfer entropy did not always reveal statistically significant results, primarily due to the limited number of years for which data was available, which impacted the transfer entropy calculations. In terms of the simulation of cascading failure in shareholding networks, the country based measurements were affected due to the fact that government interventions and other mitigating factors that may play a part during cascading failures were not modelled explicitly.  Despite these limitations, the presented work showed enough evidence to suggest that co-patenting networks and shareholding networks co-evolve among ensembles of firms which interact with each other, and that the emergence of patent links at a particular point in time could be an indicator for the corresponding emergence of shareholding links in coming years.

Future work may seek to consider other types of interactions and links between firms, such as supply chain interactions or joint ventures, and see how these interactions co-evolve or how one type of interaction could be a catalyst for other types of interactions to occur in the coming years. Again, the information theoretic measures that we have used in this work, such as mutual information and transfer trophy could play a vital part in measuring shared information and causality in such experiments. In particular, the types of non-equity interfirm interactions discussed in Grandori and Soda~\cite{grandori1995inter},      Ness and Haughland ~\cite{ness2005evolution}      and Ozman~\cite{ozman2009inter} could be considered in analysing the co-evolution of interfirm networks. An integrated cascading failure model could then be developed, which looks at not only cascading failure in shareholding networks, but  also such failures spreading through many types of links in firms. Such an integrated cascading failure model, which considers multiple types of relationships between firms, employs information theoretic measures to understand interdependency and causality between them, and simulates cascading failure based on multiple relevant parameters, could provide several important clues about the evolution, characteristics, and resilience of economic systems around the world.

\bibliographystyle{IEEEtran}
\bibliography{access}

%/EOD

\end{document}